# *Theory of Linear Chains of Metamaterial/Plasmonic Particles as Sub-Diffraction Optical Nanotransmission Lines*

Andrea Alù, and Nader Engheta

*University of Pennsylvania, Dept. of Electrical and Systems Engineering, Philadelphia, PA 19104, U.S.A.*

*Abstract*

Here we discuss the theory and analyze in detail the guidance properties of linear arrays of metamaterial/plasmonic small particles as nano-scale optical nanotransmission lines, including the effect of material loss. Under the assumption of dipolar approximation for each particle, which is shown to be accurate in the geometry of interest here, we develop closed-form analytical expressions for the eigen-modal dispersion in such arrays. With the material loss included, the conditions for minimal absorption and maximum bandwidth are derived analytically by studying the properties of such dispersion relations. Numerical examples with realistic materials including their ohmic absorption and frequency dispersion are presented. The analytical properties discussed here also provide some further physical insights into the mechanisms underlying the sub-diffraction guidance in such arrays and their fundamental physical limits. Possibility of guiding beams with sub-wavelength lateral confinement and reasonably low decay is discussed offering the possible use of this technique at microwave, infrared and optical frequencies. Interpretation of these results in terms of nanocircuit concepts is presented, and possible extension to 2-D and 3-D nanotrasnsmission line optical metamaterials is also foreseen.

*1.     Introduction*

One of the main technological challenges of today deals with squeezing the dimensions of electronic components and/or raising their frequency of operation in order to have more compact and faster communications and computational abilities. For these purposes, the interest in designing efficient, sub-wavelength guiding structures at optical frequencies has increased in recent years. As is well known, the material conduction at infrared and higher frequencies changes its usual properties [1] that are widely exploited in microwaves, and the guiding mechanisms relying on the total reflection at a highly conducting boundary are no longer easily possible at these high frequency regimes.

An interesting alternative, however, may be found in the plasmonic resonances (see, e.g., [2]-[3]), typical of noble metals, polar dielectrics, certain semiconductors [4], and metamaterials, that arise when these materials are interfaced with conventional media [5]. The negative real part of permittivity of plasmonic materials at THz, infrared and optical frequencies, combined with reasonably low losses, is indeed responsible for several anomalous electromagnetic properties, which have been studied since the initial development of the electromagnetic theory [6]. It is well known, for instance, how an isolated plasmonic particle can support a sub-wavelength "quasi-static resonance" (i.e., local surface plasmon resonance), under suitable conditions on its geometrical and electromagnetic parameters.

As for applications of such resonances to the problem of guiding of waves at these high frequencies, chains of such plasmonic particles have been proposed as possible sub-diffractive waveguides in the infrared and optical regime, since the coupling among the individual resonances of each nanoparticle may guide a beam with sub-wavelength lateral confinement and reasonably low attenuation factor [7]-[28]. Under suitable conditions, it has been shown theoretically and experimentally by various groups how these arrays may guide the energy with a modal cross section smaller than the free-space wavelength and for a reasonable distance. In many cases reported in the literature, however, the theoretical analyses have been conducted in the "quasi-static limit" and/or for arrays under some relevant approximations. Moreover, due to convergence problems relatively less attention has been paid to the influence of material losses and frequency dispersion of the modes guided by these structures – aspects that we fully consider in the present paper. Such anomalous guidance along plasmonic particle arrays has been extensively simulated numerically or proven experimentally by various groups. It is of great interest to highlight a proper full-wave theoretical interpretation of the results, particularly the role of material loss on modal dispersion.

Our group has been interested in exploring the theoretical conditions under which this technique may be successfully utilized in the case of arrays of resonant, naturally plasmonic or artificially-made metamaterial nanoparticles, and to investigate in more details the properties of such guidance with respect to the parameters involved in this phenomenon [27]-[28], including the loss mechanism. Why more analysis for such a problem? The goal of this work is to fully review the guiding characteristics of an infinite array of particles from a general full-wave analytical point of view (even including material loss), under the only assumption of the dipolar contribution of each particle to the interaction among the particles, which, as we show in the following, is necessary for the guidance of low-attenuating beams. In this case we are able to find the fundamental physical limits of this phenomenon that would allow us to predict the general behavior of electromagnetic interaction among these particles and to obtain further physical insights. In particular, this general closed-form dispersion relation for the modes in this setup may provide us with interesting conditions on the required properties for minimal absorption or radiation damping and better robustness or higher bandwidth for these waveguides. The results obtained here may therefore offer some fundamental lower limits on the guidance properties of such plasmonic arrays.

The paper is organized as follows: in the next section the general conditions and properties of the guidance of sub-diffraction beams along plasmonic arrays are derived assuming a generic model for the polarizability of each particle composing the chain. In the following section, these theoretical results are applied to realistic plasmonic particles, analyzing also some numerical examples that confirm the theory presented here and providing physical interpretation of these results.

All over the paper, a monochromatic $e^{-i\omega t}$ time dependence is assumed.

## 2. *Modal properties of chains of polarizable particles*

In the limit in which a sub-wavelength low-loss nanoparticle is close to its dipolar resonance, its near field as well as its far field is strongly dominated by the corresponding dipolar contribution. In this case, the particle's presence, as seen by an observer placed anywhere outside its volume, may be interpreted as a corresponding dipole of amplitude $\mathbf{p} = \alpha_{ee}\mathbf{E_0}$, where $\mathbf{E_0}$ is the averaged local external electric field applied on the particle (which may be considered uniform due to the small electrical extent of the particle volume) and $\alpha_{ee}$ is its electric polarizability, here considered to be isotropic for sake of simplicity[1]. In the following, we assume the particles to operate sufficiently close to the resonance of their electric dipole moment, and thus we can safely consider each particle to be represented by $\mathbf{p}$, bearing in mind that for plasmonic particles with a negative real part of their permittivity this situation is not uncommon [2]. An analogous situation is found when magnetic dipoles or metamaterial particles that may support electric and/or magnetic resonances [29] are considered, and therefore analogous analyses may be conducted in such cases. The case in which both magnetic and electric resonances are excited simultaneously in a chain problem has been investigated in [24]. In the following, without loss of generality we concentrate on the electric case.

---

[1] The case of anisotropic particles may follow the same theoretical analysis presented here as long as the polarization of the field on each particle is aligned with one of the principal axes of anisotropy. This is often the case when non-symmetric geometries are considered, like nano-wires or short dipoles. A generalization of this analysis to arbitrary anisotropy of the particles is beyond the scope of this manuscript.

Although it is well known that the far field from a sub-wavelength particle is well described by this dipolar approximation [2]-[3], it should be noted that the assumption that the particle is near its dipolar resonance, corresponding to the condition for its polarizability to fulfill $\text{Re}\left[\alpha_{ee}^{-1}\right] \simeq 0$, ensures that even its near field is dominated by this dipolar term[2]. This implies that when considering the mutual interaction among particles near their dipolar resonance, as is the case here, it is sufficient to consider their dipolar term even when their center-to-center distance is very small. As we show in the following, this requirement of being sufficiently close to their resonances is necessary for the guidance of low-attenuating modes, and therefore the assumption that the dipolar contribution dominates is not restrictive in the following analysis.

*a) Polarizability of an isolated particle*

The electric polarizability of a generic isolated particle may be expressed in the following closed form in terms of its Mie scattering coefficient $c_1^{TM}$ by comparing the Mie $TM^r$ spherical harmonic for $n=1$ with the dipolar field generated by $\mathbf{p} = \alpha_{ee}\mathbf{E_0}$:

---

[2]Even though the higher-order multipoles of order $n$ grow faster than the dipolar one as $r \to 0$, i.e., as $r^{-2n-1}$, their amplitude on the surface of the particle is proportional to $a^{2n+1}$ [29], with $a$ being its averaged radius, ensuring that for any $r > a$, regardless of how small $a$ is, the resonant dipolar field distribution is dominant over any other non-resonant higher-order multipole contribution.

$$\alpha_{ee} = -\frac{6\pi i \varepsilon_0 c_1^{TM}}{k_0^3}, \tag{1}$$

following the notation of [29]. Here $k_0 = \omega\sqrt{\varepsilon_0 \mu_0} = 2\pi/\lambda_0$ is the background (i.e., host) wavenumber, with $\varepsilon_0$, $\mu_0$ and $\lambda_0$ as the background permittivity, permeability and wavelength, respectively, and:

$$c_1^{TM} = -\frac{U_1^{TM}}{U_1^{TM} + iV_1^{TM}}, \tag{2}$$

where $U_1^{TM}$ and $V_1^{TM}$ are real functions when the particle is lossless and the background is transparent (i.e., $k_0 \in \mathbb{R}$). As an example, the general expressions of $U_1^{TM}$ and $V_1^{TM}$ for a two-layered core-shell sphere are reported in [29].

It is well known that the dipolar resonance of a particle occurs when $V_1^{TM} = 0$, implying $c_1^{TM} = -1$ (which is the maximum absolute value that this coefficient may yield) and $\alpha_{ee} = \frac{6\pi i \varepsilon_0}{k_0^3}$, which becomes a purely imaginary quantity (as a symptom of the resonant phenomenon, the induced dipole moment is indeed 90º out of phase with the impinging excitation and thus the power extracted from the external field, $\mathbf{P} = -\frac{i\omega \mathbf{p}^*}{2} \cdot \mathbf{E}_0$ is purely real).

Note that the expressions (1) and (2), with the assumption of lossless particles and transparent background, imply that $\text{Re}\left[\left(c_1^{TM}\right)^{-1}\right] = -1$ independent of the particle geometry, and therefore:

$$\text{Im}\left[\alpha_{ee}^{-1}\right] = -\frac{k_0^3}{6\pi\varepsilon_0}. \tag{3}$$

This above condition is totally independent of the particle design, and it is consistent with the radiation condition [3] and with the power conservation issues [8]. The shape of the particle, on the other hand, affects the real part of this quantity, and therefore the resonance of a generic isolated particle in the lossless limit may be simply indicated by the condition $\text{Re}\left[\alpha_{ee}^{-1}\right]=0$, since the imaginary part of this quantity is governed by (3). When low material losses are present, they add an additional negative contribution to the right hand side of (3), which we indicate in the following as $-\alpha_{loss}^{-1}$, but they do not affect sensibly this resonance condition.

*b) Modal dispersion for a linear chain of polarizable particles*

Consider now the case of an infinite array of such polarizable dipoles, as depicted in Fig. 1, located along the $x$ axis at the location $x=Nd$ with $N$ being any positive or negative integer. Let $\mathbf{p}_N$ be the dipole moment for the $N$ th particle. Suppose now that the particle at $x=0$ is excited with a given linearly polarized electric field inducing the dipole moment $\mathbf{p}_0$. Due to the linearity of the problem and the symmetry around the $x$ axis, without loss of generality we may split the problem into the two cases of longitudinal excitation, for which $\mathbf{p}_0 \parallel \hat{\mathbf{x}}$, as in Fig. 1a, and transverse polarization, for which $\mathbf{p}_0 \perp \hat{\mathbf{x}}$, as in Fig. 1b. In both cases, as can be easily verified, due to the electric field distribution of the dipolar pattern, the other particles are also polarized in the same direction:[3] $\mathbf{p}_N \parallel \mathbf{p}_0 \ \forall N$. In the following, we look for the conditions under which such linearly polarized propagating modes in the form $\mathbf{p}_N = \mathbf{p}_0 e^{i\beta Nd}$ may be self-sustained by such a chain with a small, or even a null, attenuation factor (i.e., $\text{Im}[\beta] \simeq 0$).

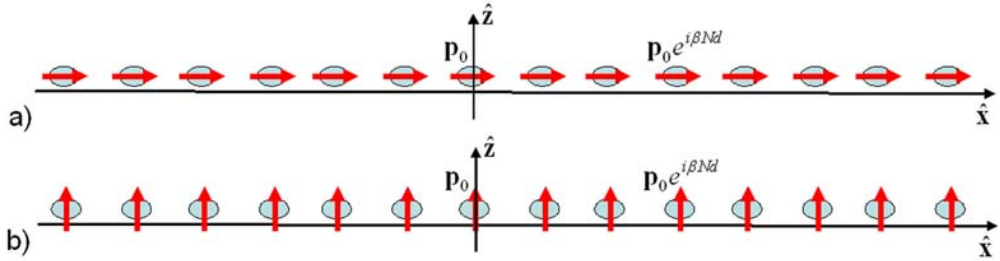

Figure 1 – Geometry of the problem: an array of metamaterial/plasmonic polarizable nanoparticles supporting a longitudinal (a) or a transverse (b) mode

In the fully dynamic case the dispersion relations for these modes may be found by supposing that such a modal distribution is somehow established along the chain and by imposing that the electric field induced at the location $x = 0$ in the absence of the particle at $x = 0$, due to the infinite chain of dipoles (removing the one at $x = 0$), excites the dipole moment $\mathbf{p}_0$ at the particle seated at $x = 0$, as already done in [8], [13], [20], [22]-[28]. In our notation this may be written in the following convenient form for the two polarizations:

---

[3] Notice that for a generic orientation of $\mathbf{p}_0$ the dipole moments induced on the other particles are not necessarily oriented in the direction parallel to $\mathbf{p}_0$. It is therefore convenient without loss of generality to decompose the problem in these two fundamental polarizations, which support linearly polarized propagation along the chain. Arbitrary polarization of the guided modes is considered as a special case in the next section.

$$L: 6\sum_{N=1}^{\infty}\left[N^{-3}\overline{d}^{-3}\cos\left(N\overline{\beta}\overline{d}\right)e^{iN\overline{d}}\left(1-iN\overline{d}\right)\right]=\overline{\alpha}_{ee}^{-1}$$

$$T: -3\sum_{N=1}^{\infty}\left[N^{-3}\overline{d}^{-3}\cos\left(N\overline{\beta}\overline{d}\right)e^{iN\overline{d}}\left(1-iN\overline{d}-N^{2}\overline{d}^{2}\right)\right]=\overline{\alpha}_{ee}^{-1}$$

(4)

where $\overline{d}=k_0 d$, $\overline{\beta}=\beta/k_0$, $\overline{\alpha}_{ee}=k_0^3\alpha_{ee}/(6\pi\varepsilon_0)$. In this way all the quantities involved in Eq. (4) are dimensionless and the system is frequency invariant. We notice how the dispersion equations are even with $\beta$, as expected, and periodic in $\beta$ with principal period $-\pi/\overline{d}<\overline{\beta}<\pi/\overline{d}$, due to the inherent periodicity of the Floquet modes of the chain.

The problem has been elegantly reduced in (4) to two dispersion equations involving three quantities: the normalized spacing $\overline{d}$, related to the geometrical properties of the chain, the normalized inverse polarizability $\overline{\alpha}_{ee}$, related to the particle properties, and the normalized guided wave number $\overline{\beta}$, which is the unknown of the problem and the quantity of interest.

For the two polarizations Eq. (4) is a complex equation, with the complication that the series on the left-hand side are slowly convergent with $N$. Moreover, these summations diverge for complex solutions of $\beta$, as noticed in [20]. This is because strictly speaking the supported complex mode would be growing exponentially at $+\infty$ or $-\infty$, depending on the chosen sign of $\overline{\beta}$, in a way faster than the decay of the dipolar fields. Clearly this problem is not present when the chain is fed at a finite $x$, as in any physically realizable system, but it arises under the hypothesis of a modal distribution established at infinity. This divergence problem might explain why in the technical literature less attention has been paid to the theoretical full-wave analysis of the behavior of such eigenmodes propagating along infinite chains of plasmonic particles when material losses are considered or when the modes radiate in the background medium, for which case the solutions for $\bar{\beta}$ would necessarily be complex.

This problem can be overcome by regularizing (4). This is done by supposing that the dispersion equations admit real solutions, and then applying an analytical continuation in the complex plane. In this case, it is convenient to rewrite the equations in terms of polylogarithm functions $Li_N(.)$:

$$L: 3\bar{d}^{-3}\left[f_3(\bar{\beta},\bar{d}) - i\bar{d}\,f_2(\bar{\beta},\bar{d})\right] = \bar{\alpha}_{ee}^{-1}$$
$$T: -\frac{3}{2}\bar{d}^{-3}\left[f_3(\bar{\beta},\bar{d}) - i\bar{d}\,f_2(\bar{\beta},\bar{d}) - \bar{d}^2 f_1(\bar{\beta},\bar{d})\right] = \bar{\alpha}_{ee}^{-1}$$  (5)

with $f_N(\bar{\beta},\bar{d}) = Li_N\left(e^{i(\bar{\beta}+1)\bar{d}}\right) + Li_N\left(e^{-i(\bar{\beta}-1)\bar{d}}\right)$.

Polylogarithm's standard definition, as given in [30], is $Li_N(z) = \sum_{k=1}^{\infty} \frac{z^k}{k^N}$ and this function is strictly convergent only for $|z| \leq 1$. However, as first noted by Euler, the functions can be analytically continued in the complex plane when written in the integral iterative form $Li_N(z) = \int_0^z \frac{Li_{N-1}(t)}{t} dt$, with $Li_1(z) = -\ln(1-z)$ [30]. In this way the dispersion relations (5) are convergent for real and complex values of $\bar{\beta}$ and, due to their numerous analytical and recursive properties, their evaluation is very efficient and it can be performed instantaneously with standard mathematical software (e.g., [31]). This eliminates any problem in the evaluation of real and complex poles for the supported modes of the chain, and it represents a complete general expression for the problem at hand. The use of polylogarithm functions for expressing, in closed form, the dispersion of guided modes along such linear chains has also been proposed in recent contributions [25]-[26].

It is interesting to underline that formulation (5) is valid for any complex value of the parameters coming into play (i.e., when material losses are introduced or when leaky radiating modes are studied). The properties of polylogarithms and associated functions may help in understanding the intrinsic properties of this configuration (including the material loss) when utilized to guide as well as to radiate energy, as we show in the following.

It may also be noted that the analytical continuation introduced here, which allows avoiding the divergence problem present in (4), may yield complex solutions that are not strictly physical, in the sense that they may diverge at $\pm\infty$. In real systems, since the structure is finite and fed at a specific point, this problem does not constitute an issue, and the validity of such solutions follows the same mathematical discussion justifying improper leaky-wave solutions in any guiding structure, or low-attenuating modes in lossy waveguides [32]: they dominate the steepest-descent approximation in specific angular regions of the visible spectrum and, even if not representing proper physical solutions for the modes of the structure, they constitute effective descriptions of the near-field distribution at their resonance. We do not discuss further this point, since it has been widely discussed in the technical literature (see, e.g., [32]). It should be noted, however, that limiting this formulation to the case of real $\bar{\beta}$, Eq. (5) is consistent with (4), with the accelerated series expression derived in [33] or with the approximated closed-form solution of this series reported in [22]-[24].

*c) Analytical properties of the dispersion equations*

Let us consider now the case of interest here, i.e., a transparent background with $k_0 \in \mathbb{R}$ (the cases of a lossy and/or ε-negative background are also part of the analysis of the previous paragraph, but they will be analyzed elsewhere). In this case, real solutions of $\bar{\beta}$ for Eqs. (4)-(5) would correspond to propagating modes along the chain that do not radiate energy and do not decay. By extracting the imaginary part of Eq. (5) under the hypothesis of $\bar{\beta}$ being a real quantity, and noticing that [30]:

$$\begin{cases} Li_1(e^{i\theta}) = Cl_1(\theta) + i\frac{(\pi - \theta)}{2} \\ Li_2(e^{i\theta}) = \frac{\pi^2}{6} - \frac{\theta(2\pi - \theta)}{4} + i\,Cl_2(\theta) \qquad 0 \leq \theta \leq 2\pi, \\ Li_3(e^{i\theta}) = Cl_3(\theta) + i\frac{\theta(\pi - \theta)(2\pi - \theta)}{12} \end{cases} \qquad (6)$$

where $Cl_N(\theta)$ are the Clausen's functions [30], which are real for real argument and whose numerical evaluation as Clausen integrals is immediately available with any calculator (these functions are also tabulated in many sources, e.g., [34]), we may get interesting closed-form relations. It may be noted in particular that the identity $Cl_1(\theta) = -\ln[2\sin(\theta/2)]$ holds.

In the principal period $|\bar{\beta}| < \pi/\bar{d}$, which is the only physically relevant period (the higher-order Floquet modes can be all reduced to this principal period, due to

the discreteness of the system), (6) and the imaginary part of (5) yields the relations necessary for having a real-valued propagating factor:

$$\begin{cases} L: \operatorname{Im}\left[\bar{\alpha}_{ee}^{-1}\right] = -1 \\ T: \operatorname{Im}\left[\bar{\alpha}_{ee}^{-1}\right] = -1 \end{cases} \quad \text{for } 1 < |\bar{\beta}| < \pi/\bar{d}$$

$$\begin{cases} L: \operatorname{Im}\left[\bar{\alpha}_{ee}^{-1}\right] = -1 + 3\pi\left(1-\bar{\beta}^2\right)/\left(2\bar{d}\right) \\ T: \operatorname{Im}\left[\bar{\alpha}_{ee}^{-1}\right] = -1 + 3\pi\left(1+\bar{\beta}^2\right)/\left(4\bar{d}\right) \end{cases} \quad \text{for } 0 < |\bar{\beta}| < 1 \quad . \quad (7)$$

Eq. (7) represents an interesting result, consistent with what is found in [8] for the longitudinal polarization. These closed form expressions in fact ensure the power conservation issue: when $|\bar{\beta}| > 1$ the guided mode is a slow wave, implying that the interference of the dipolar fields of each particle is destructive at any visible angle in the far zone, leading to the propagation of a (non-radiating) guided surface wave along the chain. At a sufficient distance from the chain, for which only the dominant first-order Bloch mode comes into play (the higher-order Bloch modes are all evanescent in this situation), the field distribution may be evaluated by assuming the presence of an averaged current line along the $x$ axis with amplitude $-i\omega\mathbf{p}_0 e^{i\beta z}/d$, leading to a decay in the radial direction in this guided mode regime, as [35]:

$$\begin{cases} L: \quad K_1\left(\sqrt{\beta^2 - k_0^2}\sqrt{y^2 + z^2}\right) \\ T: \quad K_2\left(\sqrt{\beta^2 - k_0^2}\sqrt{y^2 + z^2}\right) \end{cases}, \quad (8)$$

for the two polarizations, with $K_n(.)$ being the modified cylindrical Bessel function of order $n$. In this case, real solutions of $\beta$ for (5) are available along lossless chains, since the condition $\operatorname{Im}\left[\bar{\alpha}_{ee}^{-1}\right] = -1$ is consistent with the physical requirement (3) for lossless particles. Physically, guided modes may indeed be supported by a lossless structure provided that their $|\beta| > k_0$, in order to ensure that the mode does not radiate power in free space and neither suffers of material absorption. This concept is elegantly summarized in Eq. (7) for the two polarizations.

This situation, which is of interest for the present paper, is possible only for $\overline{d} < \pi$, as clearly seen from the condition of validity of (7) that is $|\overline{\beta}| < \pi/\overline{d}$. This condition on the spacing between the particles, for which $d < \lambda_0/2$, represents a first fundamental limit for the guiding properties of such dipole chains. We note also that the guided beam is weakly guided for chains with spacing factor close to this limit, since when $\overline{d} \simeq \pi$ the guided modes have necessarily $\overline{\beta} \simeq 1$, which implies a mode not really confined near the particles, as shown by Eq. (8). This implies that a good guidance is possible only for particles sufficiently packed, and $|\overline{\beta}| < \pi/\overline{d}$ represents quantitatively this limit on the geometry of the chain.

When $|\overline{\beta}| < 1$, on the other hand, a positive term is added to the right hand side of the equations in (7). In this case, in fact, the interference of the dipolar patterns may add up constructively, producing a leaky-wave mode. The positive term added to the right hand side of (7) may be justified equating the averaged real power radiated by each particle (which is non zero only when $|\overline{\beta}| < 1$) to the power extracted by each dipole, as shown for the longitudinal case in [8].

In the two polarizations, working in the far field where we can simplify the problem by assuming again the presence of a current line of averaged amplitude $-i\omega \mathbf{p}_0 e^{i\beta z}/d$, we get for the following exact result for the averaged power radiated by each dipole in the radial direction:

$$\begin{cases} L: & P_{rad} = \dfrac{\omega|\mathbf{p}_0|^2\left(k_0^2-\beta^2\right)}{8d\varepsilon_0} \\ T: & P_{rad} = \dfrac{\omega|\mathbf{p}_0|^2\left(k_0^2+\beta^2\right)}{16d\varepsilon_0} \end{cases} \qquad (9)$$

which is valid only when $\beta < k_0$ (as already mentioned the radiated power is zero for $\beta > k_0$). After the due normalizations, the imaginary part of the left-hand side of (5) takes exactly into account this radiated power, adding accordingly the positive term on the right-hand side of Eq. (7) in this leaky-wave regime. Eq. (7), therefore, also describes analytically the radiation properties of such chains in their leaky-wave modal operation.

For having real solutions for $\bar{\beta}$ in this case, Eq. (7) requires the use of particles (and/or the host) with active (i.e., gain) materials, whose polarizability compensates for the radiation losses evaluated in (9). In this case the chain would of course act as a nano-leaky-wave antenna, rather than as a waveguide, which may have different interesting applications, not discussed in the present paper.

It should be noted that for the longitudinal case Eq. (7)-(9) have a smooth transition at $|\bar{\beta}|=1$, but in the transverse polarization there is a discontinuity between the two intervals. As we show in the following, the dispersion features in the two polarizations are in fact different, and only the longitudinal case allows a continuous transition from the surface-wave propagation to the leaky-wave mode of operation. This is again reflected in the properties of Eq. (7).

Remaining in the lossless limit and seeking the non-attenuating guided modes under the necessary conditions $\bar{d} < \pi$ and $1 < |\bar{\beta}| < \pi/\bar{d}$, the real part of (5) determines the guiding properties of these modes as a function of the particle properties, fully determined by $\text{Re}\left[\bar{\alpha}_{ee}^{-1}\right]$ (it is interesting to note how the imaginary part of the inverse polarizability is related to power conversation issues, represented by (7), whereas its real part determines the guidance properties of the chain). Using (6) in this case we can write:

$$L: \text{Re}\left[\bar{\alpha}_{ee}^{-1}\right] = 3\bar{d}^{-3}\left[g_3\left(\bar{\beta},\bar{d}\right) + \bar{d}\, g_2\left(\bar{\beta},\bar{d}\right)\right]$$
$$T: \text{Re}\left[\bar{\alpha}_{ee}^{-1}\right] = -\frac{3}{2}\bar{d}^{-3}\left[g_3\left(\bar{\beta},\bar{d}\right) + \bar{d}\, g_2\left(\bar{\beta},\bar{d}\right) - \bar{d}^2 g_1\left(\bar{\beta},\bar{d}\right)\right] \quad (10)$$

with $g_N\left(\bar{\beta},\bar{d}\right) = Cl_N\left[\left(\bar{\beta}+1\right)\bar{d}\right] + Cl_N\left[\left(-\bar{\beta}+1\right)\bar{d}\right]$.

The dispersion properties for the two polarizations are indeed very distinct, and they are analyzed in the following paragraphs in the guided region $1 < |\bar{\beta}| < \pi/\bar{d}$. Owing to the evenness of these dispersions, without loss of generality we focus our attention on the region of positive phase velocities, i.e., $1 < \bar{\beta} < \pi/\bar{d}$.

*d) Longitudinally-polarized modes*

In the longitudinal case, taking the derivative of the right hand side of (10) and considering the properties of the integral definition of Clausen's function

$$\frac{\partial Cl_N(\theta)}{\partial \theta} = (-1)^N Cl_{N-1}(\theta), \quad (11)$$

we find that:

$$L: \frac{\partial \operatorname{Re}\left[\bar{\alpha}_{ee}^{-1}\right]}{\partial \bar{\beta}} < 0 \qquad \forall \bar{d}. \qquad (12)$$

This implies the important result that the region of guidance of such chains in the longitudinal mode is determined by the limiting values of $\operatorname{Re}\left[\bar{\alpha}_{ee}^{-1}\right]$, as given in (10), calculated at the extremes of the interval $1 < \bar{\beta} < \pi/\bar{d}$. Applying (6) and the identities $Li_2(1) = \pi^2/6$ and $Li_3(1) = \xi(3)$, with $\xi(.)$ being the Riemann zeta function, this range may be written as:

$$L: 6\left[Cl_3(\bar{d}+\pi) + \bar{d}\, Cl_2(\bar{d}+\pi)\right] < \bar{d}^3 \operatorname{Re}\left[\bar{\alpha}_{ee}^{-1}\right] < 3\left[\xi(3) + Cl_3(2\bar{d}) + \bar{d}\, Cl_2(2\bar{d})\right]$$
(13)

This represents the closed-form expression, as a general result, for the range of polarizabilities capable of supporting guided longitudinal modes in a chain of lossless polarizable particles with a given spacing. In other words, seeking for guided modes in longitudinal linear polarization, the particles composing the chain should be designed to have the proper inverse polarizability falling in this range, depending on the particle separation.

Fig. 2 reports this guidance region in the plot of $\bar{d}$ versus $\bar{d}^3 \operatorname{Re}\left[\bar{\alpha}_{ee}^{-1}\right]$. The red dotted line corresponds to the locus $\bar{\beta} = 1$, which is the border between the guided propagation (below the line) and the leaky-wave propagation (above the line), whereas the black solid line is the locus where $\bar{\beta} = \pi/\bar{d}$, below which complex evanescent Floquet modes with $\operatorname{Re}\left[\bar{\beta}\right] = \pi/\bar{d}$ are supported.

The region below the black line, in which the phases of two neighboring particles are opposite to each other and the mode is therefore non-radiating, rapidly attenuated and totally reflected at the entrance of the chain, represents the stop-band region of this configuration. It should be noted how the stop-band for this chain may arise at frequencies for which its periodicity is much smaller than the background wavelength $\lambda_0$. Usually band-gap structures are characterized by a periodicity of the order of $\lambda_0/2$, for which Bragg reflection arises (see, e.g., [36]). The condition for entering the stop-band region in this case, instead, is that the chain spacing is $\lambda_g/2$, which $\lambda_g = 2\pi/\beta$ is much smaller than $\lambda_0$ for sufficiently packed particles. This implies that a sub-wavelength plasmonic chain may support a first band-gap in the long-wavelength regime, due to the anomalous slow-wave mode supported by such chains. This phenomenon is not discussed further in the present paper and it will be the subject of future investigations for some other interesting potential applications.

In the region between the two lines, longitudinal guided modes with no attenuation (in the limit of no losses we are considering now) are supported, with $\bar{\beta}$ monotonically increasing with a decrease of $\text{Re}\left[\bar{\alpha}_{ee}^{-1}\right]$, consistent with (12). You may notice how for small $\bar{d}$ the guidance region rapidly widens up (a factor of $\bar{d}^3$ is also normalizing the vertical axis), being centered around the resonance condition for the isolated particle $\text{Re}\left[\bar{\alpha}_{ee}^{-1}\right] = 0$. In any case, for a guided mode to be supported the requirement of particles near their resonance remains necessary, and it is physically understandable since non-resonant nanoparticles offer weak induced dipole moments and low scattering, and therefore their interaction in the chain would not be sufficient to self sustain a propagating mode. As more widely

-19-

discussed in the next section, in fact, even though the guidance region widens up around the point $\mathrm{Re}[\bar{\alpha}_{ee}^{-1}] = 0$ when $\bar{d} \to 0$ as $\bar{d}^{-3}$, the resonant bandwidth of each particle necessarily narrows down as $(k_0 a)^{-3}$ [37], with $a$ being its averaged linear dimension. These two factors compensate each other, as it is discussed in the next section.

The transition between the surface-wave and leaky-wave operations is smooth for this polarization: increasing the particle inverse polarizability, the mode becomes weakly guided and at some point starts radiating energy away, scanning the beam angle with a variation of $\mathrm{Re}[\bar{\beta}]$ and having a high directivity of the beam if $\mathrm{Im}[\bar{\beta}]$ is sufficiently low, as in any leaky wave antenna.

When the distance between particles is increased, the required polarizability of each particle in the chain for having a guided mode narrows down in a smaller range around the resonance, since the particles are required to be closer to their individual resonance to support a guided mode. However, the presence of the other scatterers shifts the resonance condition, and the guidance region for a given $\bar{d}$ is not necessarily centered around $\mathrm{Re}[\bar{\alpha}_{ee}^{-1}] = 0$, even though it remains close to it. Indeed for $1.71 < \bar{d} < 2.95$ and $3.01 < \bar{d} < \pi$, an array of lossless particles exactly at their individual resonance, i.e., with $\mathrm{Re}[\bar{\alpha}_{ee}^{-1}] = 0$, would not guide a longitudinal mode with real $\bar{\beta}$. In this case, the proper condition on the polarizability requires it to be tuned downwards or upwards in the plot. When $\bar{d} \to \pi$ the range of required polarizabilities tends to a single specific value $\text{Re}\left[\bar{\alpha}_{ee}^{-1}\right] = 6\pi^{-3}\xi(3)$, for which $\bar{\beta} \to 1^+$. This is the only possible value of polarizability for an infinite chain of lossless particles with spacing $d = \lambda_0/2$ in order to support a resonance. For distances beyond this value no resonance (or guided waves) can be supported by such an array, consistently with the results of the previous paragraph.

Curves for different values of $\bar{\beta}$ are also reported in the plot. The region included between the green dotted line and the solid black line, for instance, are those for which $\bar{\beta} > 3$. This makes it clear that a more negative inverse polarizability in the region of guidance confines more the guided beam around the chain. Similar curves are plotted for $\bar{\beta} = 2$ and $\bar{\beta} = 1.5$, also showing the narrower limits in the spacing between the particles to support a beam concentrated around the chain.

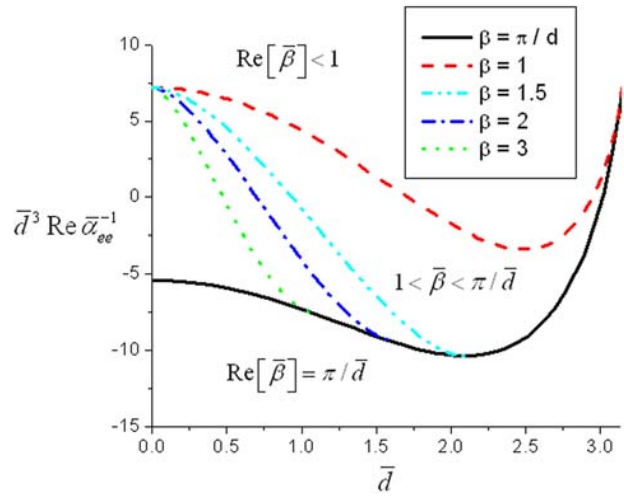

Figure 2 – Regions of guidance for the longitudinal mode (Fig. 1a). Above the red-dashed line, a longitudinal leaky-mode exists, whereas below the black solid line we enter the band-gap region of the chain in this polarization. For values of polarizabilities between the two lines guided

propagation is possible, with the other lines delimiting more stringent ranges for the higher values of $\bar{\beta}$, and therefore more confined modes.

Figure 3 reports the dispersion diagram for the real $\bar{\beta}$ versus $\text{Re}\left[\bar{\alpha}_{ee}^{-1}\right]$ for the longitudinal modes guided by an infinite array of lossless particles in the two significant cases of narrow spacing ($\bar{d} = 0.1$, Fig. 3a) and wide spacing ($\bar{d} = 0.9\pi$, Fig. 3b). The plots are compared with the nearest neighbor approximation (NNA), i.e., the approximate solution obtained using only the quasi-static dipolar field of the nearest neighboring particle, that can be obtained from Eq. (4) by truncating the summation to the first term and neglecting the imaginary part of the solution (which is vanishing for small $\bar{d}$). This approximation has often been used in the technical literature (see e.g., [9]) for the analysis of this setup. In the plots the horizontal axis is delimited by the boundaries for the particles polarizability derived from Fig. 2. You notice how in the case of narrow spacing (Fig. 3a) $\bar{\beta}$ may yield large values, limited by $\pi/\bar{d}$, implying that a sub-wavelength confinement of the mode is possible, and the NNA predicts the dispersion reasonably well, since the coupling between neighboring particles dominates. It is interesting to note that once the mode becomes sufficiently slow, it is in principle possible to confine the beam in a region of space much smaller than the wavelength of operation, going much beyond the diffraction limit. From these results it may appear that there is no lower limit on how narrow the spacing between particles, and therefore the guided cross section, may be made (the guided beam cross section decays with an increase of $\beta$, following (8)). However, this property is limited by losses, as it will be shown in the following section.

For electrically larger spacing (Fig. 3b), on the other hand, the NNA yields incorrect results. Also, the guidance is limited to a narrow range of polarizabilities, as predicted by Fig. 2 and the mode is poorly confined along the chain, since $\bar{\beta} \simeq 1$.

It is interesting to note how for particles at their individual resonance, i.e., $\text{Re}\left[\bar{\alpha}_{ee}^{-1}\right] = 0$, the NNA predicts the approximate solution $\bar{\beta}\bar{d} = \pi/2$, with a cut-off for the guided modes in this case to be at $\bar{d} = \pi/2$, very near the exact value $\bar{d} = 1.71$ determined in our earlier discussion about Fig. 2. The discrepancy between the exact solution and the NNA is more accentuated when the mode is close to the borders of the guidance region, i.e., near the leaky-wave or the band-gap regions.

As a last comment on these plots, it is worth noting how, as predicted by (12), the slope of the dispersion curve is always negative. As will be discussed in more details later, this is a consequence of the forward-wave propagation of these longitudinal modes (i.e., phase and group velocities parallel to each other).

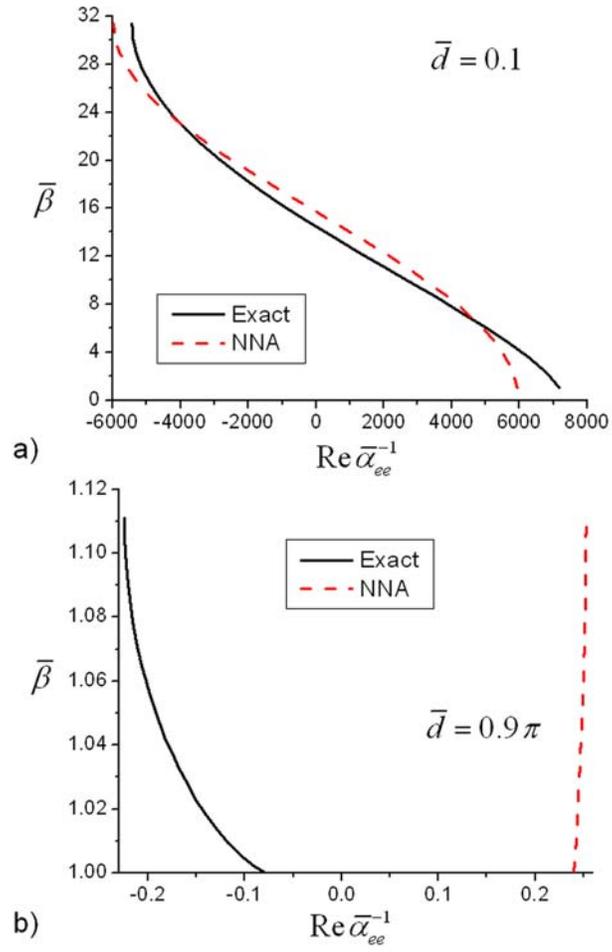

Figure 3 – Dispersion plots for the guided longitudinal modes in terms of the particle polarizability for two different values of the spacing between neighboring particles: (a) narrow spacing, $\bar{d} = 0.1$; (b) wide spacing, $\bar{d} = 0.9\pi$. NNA stands for "nearest neighbor approximation".

*e) Transversely-polarized modes*

The analysis of the case of transverse polarization is more challenging, since the dispersion curve of $\mathrm{Re}\left[\alpha_{ee}^{-1}\right]$ versus $\beta$ is not necessarily monotonic. In particular, for this polarization $\partial\mathrm{Re}\left[\bar{\alpha}_{ee}^{-1}\right]/\partial\bar{\beta}$ may flip its sign inside the range $1 < \bar{\beta} < \pi/\bar{d}$ for smaller spacing between the particles. In this situation, the chain may clearly support two distinct surface modes in a given range of polarizabilities. Taking the derivative of (10) and using the properties of Clausen's functions, it is possible to show that this coexistence of two modes is possible when $\bar{d} < \hat{d}$, with $\hat{d}$ satisfying the following transcendental equation:

$$\frac{\ln\left[2\left(1+\cos\hat{d}\right)\right]}{\hat{d}} + \tan\frac{\hat{d}}{2} = \frac{\hat{d}}{1+\cos\hat{d}}, \tag{14}$$

whose solution is $\hat{d} = 1.517$. Obviously, this property is independent of the particle polarizability.

In this situation, the two supported $\bar{\beta}$ have opposite behavior: one mode is confined around the chain, with $\bar{\beta}_{min} < \bar{\beta} < \pi/\bar{d}$ and $\partial \text{Re}\left[\bar{\alpha}_{ee}^{-1}\right]/\partial\bar{\beta} > 0$. This mode is supported in the range of polarizabilities:

$$T: \bar{d}^3 \bar{\alpha}_{min}^{-1} < \bar{d}^3 \text{Re}\left[\bar{\alpha}_{ee}^{-1}\right] < -3\left[Cl_3\left(\bar{d}+\pi\right) + \bar{d}\, Cl_2\left(\bar{d}+\pi\right) - \bar{d}^2 Cl_1\left(\bar{d}+\pi\right)\right], \tag{15}$$

whose right-hand side is obtained from (5) with $\bar{\beta} = \pi/\bar{d}$. The other supported mode has $\bar{\beta} \simeq 1$ and its value slightly increases when $\text{Re}\left[\bar{\alpha}_{ee}^{-1}\right]$ is lowered, i.e., $\partial \text{Re}\left[\bar{\alpha}_{ee}^{-1}\right]/\partial\bar{\beta} < 0$. It is supported for any $\text{Re}\left[\bar{\alpha}_{ee}^{-1}\right] > \bar{\alpha}_{min}^{-1}$, i.e., it does not have an upper cut-off limit for the inverse polarizability.

At the specific value $\text{Re}\left[\bar{\alpha}_{ee}^{-1}\right] = \bar{\alpha}_{min}^{-1}$, the two dispersion branches collapse into the same degenerate solution with $\bar{\beta} = \bar{\beta}_{min}\left(\bar{d}\right)$ and they become complex conjugates for smaller values of $\text{Re}\left[\bar{\alpha}_{ee}^{-1}\right]$, entering the complex-mode region[4].

---

[4] An analogous complex-mode regime that characterizes the guided mode spectrum of plasmonic planar slabs [38] and the anomalous power-flow properties of complex modes, which carry zero

This implies that in principle there is no upper limit for the polarizabilities required for having a guided mode with this polarization. Although when $\text{Re}\left[\bar{\alpha}_{ee}^{-1}\right]$ increases beyond the existence of the first (confined) mode, i.e., beyond the right hand limit of (15), the only supported mode is very weakly guided, close to the cut-off, and spread all over the space almost as a uniform plane wave (since $\bar{\beta} \simeq 1$).

The interest is evidently concentrated on the mode determined from condition (15), with a dispersion

$$T: \frac{\partial \text{Re}\left[\bar{\alpha}_{ee}^{-1}\right]}{\partial \bar{\beta}} > 0 \qquad \forall \bar{d}, \qquad (16)$$

contrary to the case of longitudinal polarization.

From a physical point of view, for this polarization the presence of the secondary mode, which is not even predicted by the NNA, is not surprising, since in the limit of very high values of $\text{Re}\left[\bar{\alpha}_{ee}^{-1}\right]$ the particles are just vanishingly small dielectric particles (or particles with materials similar to that of the background) weakly interacting with the field. The transverse polarization indeed includes the case of TEM plane wave propagation in the background medium with no influence from the particles, for which $\bar{\beta} = 1$. Thus it is not surprising to find a solution in this case even when the particles are vanishingly small. Decreasing the value of $\text{Re}\left[\bar{\alpha}_{ee}^{-1}\right]$, this secondary mode starts to weakly interact with the chain and its $\bar{\beta} \simeq 1$ weakly increases, up to the point in which $\bar{\beta} = \bar{\beta}_{min}$ and the two modes

---

net-power in the longitudinal direction, have been studied over the years for several geometries [39].

degenerate. This second mode is clearly not of interest in the present analysis, since it cannot be claimed that such a modal distribution is really guided by the chain[5].

The special degenerate solution $\bar{\beta} = \bar{\beta}_{min}$ for which $\text{Re}\left[\bar{\alpha}_{ee}^{-1}\right] = \bar{\alpha}_{min}^{-1}$ corresponds to the limiting case of a mode with zero group velocity (due to the superposition of the two degenerate modal solutions) but non-zero phase velocity (in fact $\bar{\beta} = \bar{\beta}_{min} > 1$). We note that at this point $\partial \text{Re}\left[\bar{\alpha}_{ee}^{-1}\right]/\partial \bar{\beta}$ is indeed zero and, for a given linear chain with assigned spacing, we can write the following equality among the three variables involved in the problem $\omega$, $\beta$ and $\text{Re}\left[\alpha_{ee}^{-1}\right]$:

$$v_g = \left.\frac{\partial \omega}{\partial \beta}\right|_{\text{Re}\left[\alpha_{ee}^{-1}\right]} = -\frac{\left.\frac{\partial \text{Re}\left[\alpha_{ee}^{-1}\right]}{\partial \beta}\right|_\omega}{\left.\frac{\partial \text{Re}\left[\alpha_{ee}^{-1}\right]}{\partial \omega}\right|_\beta}, \qquad (17)$$

where the partial derivatives are taken when fixing the variable in the pedix. Since the condition $\partial \text{Re}\left[\bar{\alpha}_{ee}^{-1}\right]/\partial \bar{\beta} = 0$ makes the numerator of (17) zero, this special situation corresponds to a zero group velocity with a non-zero phase velocity for the degenerate modes.

This phenomenon is not uncommon when dealing with guided modes in plasmonic or negative-index structures, as we have found in different configurations [41]-[44]. In this chain configuration this special solution has been investigated theoretically by Simovski et al. in [45]-[46]. We show in the following, however, how the presence of absorption in the particles affects particularly these regions of guidance for which the derivative $\partial \text{Re}\left[\bar{\alpha}_{ee}^{-1}\right]/\partial \bar{\beta}$ is particularly low, reducing the effective applicability of this region of operation. Moreover, when losses are considered, this derivative, and consequently the group velocity of the mode, can never be identically zero, even though it may become sufficiently low for a low-absorptive material.

---

[5] As an aside, a similar dual-mode of propagation is typical of spatially dispersive materials, e.g., the wire medium [40]. The 3D extension of this chain problem, in fact, shows a spatially dispersive characteristic, as will be presented elsewhere in the near future.

Fig. 4 shows the admissible range of polarizabilities for having guided modes in this polarization. The black solid line, which is the locus for which $\bar{\beta} = \pi/\bar{d}$ when $\bar{d} < \hat{d}$, represents an upper boundary for the polarizability in order to have the confined mode in the two-mode regime (with $\bar{\beta}_{min} < \bar{\beta} < \pi/\bar{d}$). Beyond this line, in fact, we enter the band-gap region for this confined mode, since its $\bar{\beta}$ is sufficiently high to have $\bar{d} = \lambda_g/2$. The red dashed line, on the other hand, provides the value of polarizability that gives $\bar{\beta} = \bar{\beta}_{min}$, i.e., it represents the value of $\bar{\alpha}_{min}^{-1}$ as a function of $\bar{d}$. Below the red line the two supported modes become complex conjugate of each other with $\text{Re}\left[\bar{\beta}\right] = \bar{\beta}_{min}$. The red line represents the dispersion curve for the zero-group velocity mode described in [45]-[46] and the value of its guided wave number $\bar{\beta}_{min}$ as a function of $\bar{d}$ is reported in Fig. 5, remaining close to unity in all the admissible range $\bar{d} < \hat{d}$.

The area between the black solid and red dashed line is therefore the required range for having a confined transverse guided wave propagating along the chain with no radiation. The green dotted line continues the lower limit of polarizabilities when $\bar{d} > \hat{d}$, still being the locus for which $\bar{\beta} = \pi / \bar{d}$. Below this line again no mode can be guided by the chain. In the region above the green and the back lines, however, the chain still in principle supports one guided mode, even though its distribution is very sparse in the outer region. As an example, the blue dash-dotted line is the locus where this poorly confined mode has $\bar{\beta} = 1.001$. Clearly, above this line the chain is not guiding any wave, since its distribution is very widespread in the surrounding space. The inverse polarizabilities in fact are far from their individual resonances and therefore the guided mode is close to the TEM plane wave propagating in the background medium, weakly affected by the presence of the chain. Similar to the longitudinal case, the cyan dash-dot-dotted line is the locus for $\bar{\beta} = 3$, and the region between it and the black solid line includes the region where $\bar{\beta} > 3$. Contrary to the previous case, here a more positive inverse polarizability within the region of guidance of the confined mode would produce a higher $\bar{\beta}$ and a more concentrated beam. This is due to the backwardness of the confined mode, as we will discuss later. Comparing the two polarizations, we note how the longitudinal and transverse modes have somewhat similar features for guidance, since the range of polarizabilities that would support a guided confined beam, forward or backward depending on the polarization, is comparable in the two cases.

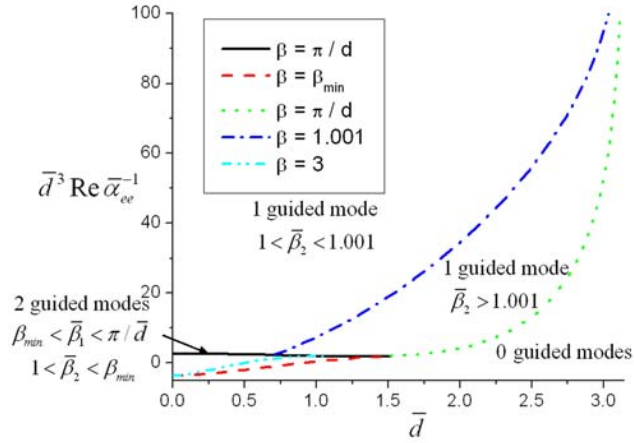

Figure 4 – Region of guidance for the transverse mode (Fig. 1b), analogous to Fig. 2. The confined mode is supported only in the region between the black (solid) and the red (dashed) lines.

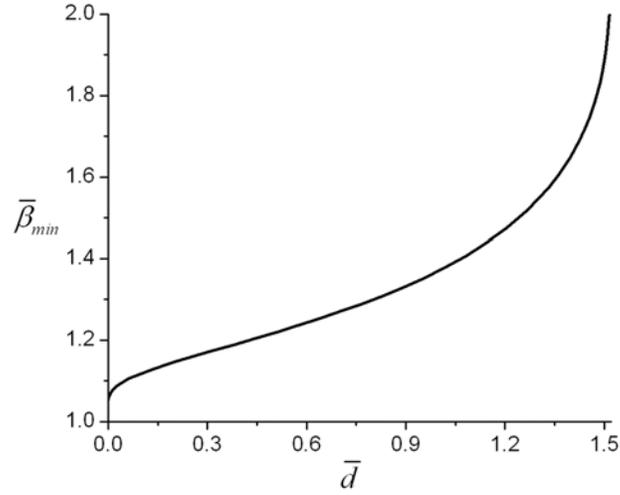

Figure 5 – Variation of $\bar{\beta}_{min}$, wave number of the zero group velocity mode, that limits the propagation of confined guided modes in the transverse polarization, as a function of $\bar{d}$

To provide an example of the dispersion for this transverse configuration, we have considered the cases reported in Fig. 6, distinguishing again between a narrow spacing ($\bar{d} = 0.1$, Fig. 6a) and a wide spacing ($\bar{d} = 0.9\pi$, Fig. 6b). In the first case, the propagation of a confined mode, similar to the longitudinal case, is shown to be possible. As predicted by (16), the slope of the curve in the confined region is opposite with respect to that of the longitudinal polarization, a consequence of the change in the direction of power flow, since, as shown in the following, this mode is backward, i.e., group and phase velocities are anti-parallel. This situation is possible provided that condition (15) is satisfied for $\bar{d} < \hat{d}$. In this case we are indeed in the region confined between the red and black lines in the plot of Fig. 4 and a second mode is visible in the region $\bar{\beta} = 1^+$. Even though this mode has in principle no upper limit for the required polarizability value, its propagation properties are clearly less appealing, being analogous to those of a plane wave propagating in the background medium and weakly interacting with the chain. The NNA predicts well the dispersion of the first mode, and it does not predict the second mode of propagation. We note the position of the zero-group velocity mode at the connection between the two curves in the plot.

For wider spacing, beyond the 2-mode region, i.e., for $\bar{d} > \hat{d}$, the only supported mode maintains a positive slope in the plot, similar to the longitudinal mode, and it is in fact a forward-wave. However, there is still no upper boundary for the inverse polarizability with $\bar{\beta}$, that rapidly approaches unity when $\text{Re}\left[\alpha_{ee}^{-1}\right]$ is increased. In this polarization, the NNA again fails completely when a wider spacing is considered and the zero-group velocity mode is not supported in this configuration.

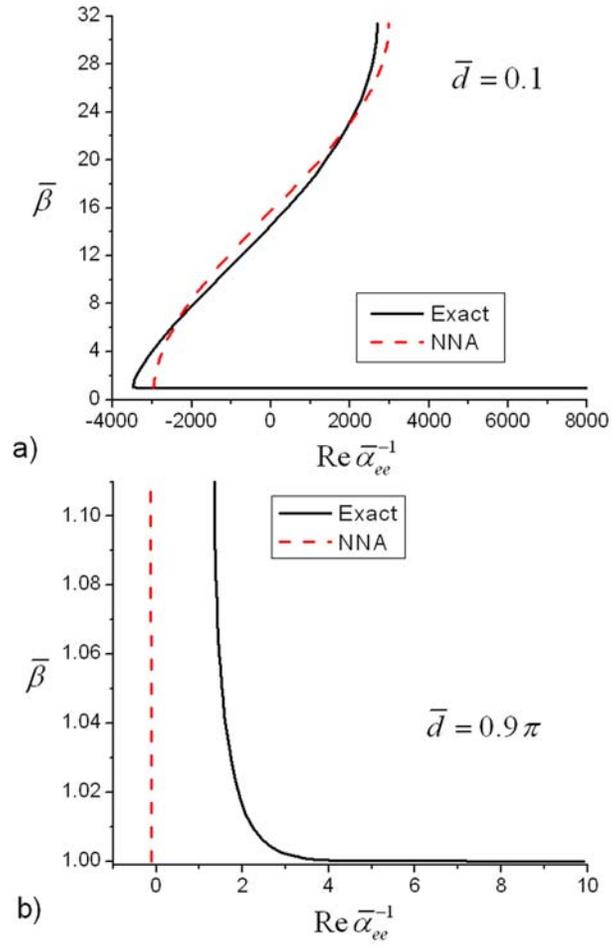

Figure 6 – Dispersion plots for the guided transverse modes varying the particles polarizability for two different values of the spacing

*f) Lossy particles*

We recall now that the analytical continuation of the dispersion relations represented by Eq. (5) also allows treating complex solutions of $\bar{\beta}$. This implies that the leaky-wave complex mode regime may be analyzed following similar steps, even though it is not of interest for the present analysis. It will be presented in a future work for different applications.

It is however of interest to consider complex solutions for $\bar{\beta}$ in the case in which small ohmic losses are added to the particles polarizability. Under this assumption, which brings the present analysis closer to a realistic model, an attenuation factor in the propagation, i.e., an imaginary part of $\bar{\beta}$, is expected. Eq. (5) in this case would indeed support complex solutions, and Eq. (4) would not properly converge. In parallel, condition (7) for having real solutions cannot be satisfied even in the region $1 < |\bar{\beta}| < \pi/\bar{d}$, since the presence of ohmic losses adds a contribution to the imaginary part of the inverse polarizability, in the form:

$$\text{Im}\left[\bar{\alpha}_{ee}^{-1}\right] = -1 - \bar{\alpha}_{loss}^{-1}, \qquad (18)$$

where $\bar{\alpha}_{loss}^{-1} > 0$ for passive particles.

For low-loss particles, of interest here, $\bar{\alpha}_{loss}^{-1}$ is small and only slightly perturbs the real solutions found in the previous analysis. In particular, the new solution satisfying (5) can be written in the form $\bar{\beta} = \bar{\beta}_r + i\bar{\beta}_i$, with $\bar{\beta}_r$ and $\bar{\beta}_i$ being real valued quantities. In the limit of small losses, perturbing (5) with the presence of a small $\bar{\alpha}_{loss}^{-1}$, expanding in Taylor series to the first order, and applying (6), we find:

$$\bar{\beta}_i = -\bar{\alpha}_{loss}^{-1} \frac{\partial \bar{\beta}_r}{\partial \text{Re}\left[\bar{\alpha}_{ee}^{-1}\right]} \qquad (19)$$

with $\bar{\beta}_r$ satisfying the unperturbed Eq. (10).

This implies that the attenuation factor of the guided modes (due to the small material losses in particles) is directly proportional to the contribution of ohmic losses to the inverse polarizability and to the derivative of the guided wave number with respect to the inverse polarizability itself, which according to (17) is inversely proportional to the modal group velocity. This is valid in both polarizations. The properties of Eq. (5) ensure that the perturbation of $\bar{\beta}_r$ due to the presence of material losses are only a second-order effect, and small ohmic losses affect mainly the attenuation factor, inducing a $\bar{\beta}_i$ proportional to $\bar{\alpha}_{loss}^{-1}$.

As a corollary to (19), we can determine the direction of power flow in the modes previously described. In fact, the regions where $\partial \bar{\beta}_r / \partial \operatorname{Re}\left[\bar{\alpha}_{ee}^{-1}\right]$ is negative (positive) are the regions where the guided mode is forward (backward). This is because under the $e^{i\beta x}$ assumption, for a positive phase velocity ($\bar{\beta}_r > 0$), a positive group velocity and power flow (i.e., a forward propagation) can be expected only when $\bar{\beta}_i > 0$, that is for a negative value of $\partial \bar{\beta}_r / \partial \operatorname{Re}\left[\bar{\alpha}_{ee}^{-1}\right]$, which happens in the longitudinal polarization and in the low-confined propagation of the transverse mode. Conversely, when $\partial \bar{\beta}_r / \partial \operatorname{Re}\left[\bar{\alpha}_{ee}^{-1}\right] > 0$, as it is for the concentrated mode in the transverse polarization, the propagation is necessarily backward, again following (19) and applying causality. Moreover, as already anticipated, the regions where $\partial \operatorname{Re}\left[\bar{\alpha}_{ee}^{-1}\right] / \partial \bar{\beta}$, and correspondingly the group velocity, is small are those more affected by the presence of ohmic losses in the particles, and those for which the attenuation of the guided mode is expected to be higher for a given material loss factor. This should be considered in applications when utilizing these chains having modes with very low (or even zero) group velocity, as in [45]-[46], since material losses are expected to strongly affect these configurations.

Using (19), it becomes straightforward to identify the more appealing regions of dispersion curves for getting a low-attenuation propagation along the chain: first of all, the sensitivity to losses increases when the spacing $\bar{d}$ is increased, as evident from the previous figures, since the derivative $\partial \bar{\beta}_r / \partial \operatorname{Re}\left[\bar{\alpha}_{ee}^{-1}\right]$ in general grows with $\bar{d}$. This is obviously related to the fact that more particles per unit length are expected to guide better the energy along the chain. Also, the region of weak guidance for closely packed particles in the transverse polarization is very weakly affected by losses (the derivative is almost zero for such a modal operation), due to the fact that the mode is very widespread around the background space and weakly interacting with the lossy particles. Near the cut-off of the modes the derivative becomes high, and we expect higher sensitivity to losses when the propagating mode is entering a cut-off region. This is consistent with the properties of any traveling-wave structure.

It is particularly interesting to apply this result in order to find the conditions of minimum absorption for the guided modes in the two polarizations, which may be useful for the design of low-attenuation waveguides and nanotransmission lines in this configuration. Based on the above discussion and following (19), these conditions are those for which the second derivative $\partial^2 \operatorname{Re}\left[\bar{\alpha}_{ee}^{-1}\right] / \partial \bar{\beta}_r^2 = 0$. Resorting again to the properties of Clausen's functions (11) and conducting some mathematical manipulations, this condition may be written for the two polarizations as:

$$L: \ln\left(2\left[\cos(\bar{d}) - \cos(\bar{\beta}_r\bar{d})\right]\right) = -\frac{\bar{d}^2 \sin\bar{d}}{\cos(\bar{d}) - \cos(\bar{\beta}_r\bar{d})}$$

$$T: \ln\left(2\left[\cos(\bar{d}) - \cos(\bar{\beta}_r\bar{d})\right]\right) = -\frac{\bar{d}^2 \sin\bar{d}}{\cos(\bar{d}) - \cos(\bar{\beta}_r\bar{d})} + \frac{\bar{d}^3}{4}\left(\sin^{-2}\frac{(\bar{\beta}_r - 1)\bar{d}}{2} + \sin^{-2}\frac{(\bar{\beta}_r + 1)\bar{d}}{2}\right)$$

(20)

For a given spacing $\bar{d}$, the optimum $\bar{\beta}$ is found by numerically solving (20). For both polarizations, this is plotted in Figure 7a after being normalized to $\bar{d}$ for plotting convenience. In the limit of $\bar{d} \to 0$, i.e., closely packed particles, the optimum $\bar{\beta}$ that ensures smaller losses interestingly tends to the value $\bar{\beta}_{opt} = \pi/(3\bar{d})$ in both polarizations. Increasing the spacing the two polarizations behave differently, as evident from the plots. If in the longitudinal polarization this value of $\bar{\beta}$ quickly converges to the regions of weakly guided modes with $\bar{\beta} \simeq 1$ (and in any case this does not admit a solution for $\bar{d} > 0.58$), the transverse polarization in this sense offers a wider range of spacing values where a concentrated mode allows achieving the minimum loss condition. Note that in both polarizations the limit of $\bar{\beta} \to 1^+$ yields $\partial\bar{\beta}/\partial\,\text{Re}\left[\bar{\alpha}_{ee}^{-1}\right] = 0$. In the transverse polarization this mode is obtainable only when the particles are removed, i.e., when $\text{Re}[\bar{\alpha}_{ee}] = 0$, which coincides with the trivial solution of a TEM plane wave propagating in the lossless background medium, whereas in the longitudinal polarization we need a specific value of the polarizability to induce a longitudinally polarized plane wave with $\bar{\beta} = 1$. This value is determined by the boundary between the guided region and the leaky-wave region for this polarization, as given by (13) and represented by the red dashed line in Fig. 2.

This analysis shows that this mode would be particularly robust to the material losses, since it is spread all over the background material. This, however, is not of interest in our present analysis, where we are concentrating on low-loss conditions for confined guided waves around the chain.

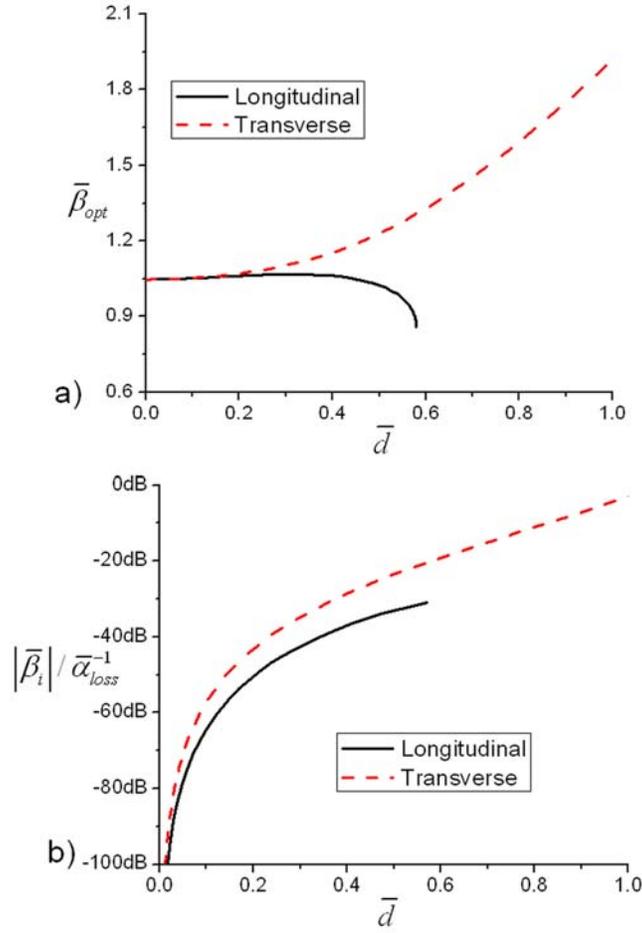

Figure 7 – (a) Value of $\bar{\beta}_{opt}$ to get the minimum attenuation loss factor in the two polarizations in terms of $\bar{d}$ ; (b) minimum $|\bar{\beta}_i|/\bar{\alpha}_{loss}^{-1}$ versus $\bar{d}$

In Fig. 7b we plot the value of $|\bar{\beta}_i|/\bar{\alpha}_{loss}^{-1}$ obtained at the optimum value $\bar{\beta}_{opt}$ given in Fig. 7a, in order to show the sensitivity of the two cases of polarizations to a given level of material losses. In this sense the longitudinal polarization offers a better performance, even though the guided beam is comparatively less confined, as Fig. 7a shows. A trade-off between these two quantities appears necessary in order to obtain an optimum design. We notice that the results in Fig. 7 represent intrinsic physical limitations to the attenuation factors of guided propagation along such chains, regardless of the nature of the particles in the chain. If a more concentrated mode is desired for a given spacing, it may be obtained at the expense of increasing the sensitivity to material loss, since the field levels at the particles' location are higher.

Once the value of $\bar{\beta}_{opt}$ is chosen for a given spacing, as obtained from the charts of Fig. 7, the optimum value of polarizability for the particles in the chain can be directly obtained from (10), and it can be used as a design tool to get the minimum attenuation factor and the desired lateral confinement of the beam around the chain.

*g) Nearest-neighbor approximation*

To conclude this section it should be pointed out that for closely-spaced particles in both polarizations the NNA can predict with a good approximation the guiding properties of the chain, as it has been shown in the previous examples, and therefore the approximate dispersion equations for small spacing may be given by:

$$\begin{aligned} L: \bar{\beta}\bar{d} &= \cos^{-1}\left[\operatorname{Re}\left[\bar{\alpha}_{ee}^{-1}\right]/\left(6\bar{d}^{-3}\right)\right] \\ T: \bar{\beta}\bar{d} &= \cos^{-1}\left[-\operatorname{Re}\left[\bar{\alpha}_{ee}^{-1}\right]/\left(3\bar{d}^{-3}\right)\right] \end{aligned} \quad (21)$$

which are obtained by truncating Eq. (4) to the first order and taking the limit for small $\bar{d}$. In this limit, applying (19), we get also:

$$L: \bar{\beta}_i = \bar{\alpha}_{loss}^{-1} \bar{d}^2 / 6$$
$$T: \bar{\beta}_i = -\bar{\alpha}_{loss}^{-1} \bar{d}^2 / 3 \quad (22)$$

confirming that in the limit of small spacing the robustness to reasonable amount of losses is high (and it is twice as stronger in the longitudinal forward case than in the transverse backward case). As we show more in details in the following section, even though the individual particle resonance is affected by losses in a way proportional to the volume of the particles, together with $\bar{\alpha}_{loss}^{-1}$, the proximity to other resonant particles increases their robustness, due to the term $\bar{d}^2$ in (22).

It should also be noted how the approximate dispersion given in (21) reveals the anomalous feature valid for closely-spaced particles, i.e., shrinking the scale of our setup, which in this case is achieved by reducing the spacing between the dipoles, increases the concentration of the guided modes around the chain, whereas usually in conventional guided-wave setups the behavior is opposite, i.e., reducing the dimensions of open slab waveguides causes the field to be widespread in the background region. This is analogous to what happens in other guiding geometries involving metamaterials or plasmonic materials, i.e., planar slabs [41] and cylindrical rods [42] that we have recently analyzed. In these structures, the mode can be guided beyond the diffraction limits, since the lateral confinement of the guided beam may be considerably reduced below the wavelength of operation when the dimension of the waveguide is reduced. The cylindrical or planar open waveguide setups may indeed be considered as limiting cases of the dipolar chains analyzed here when the spacing $d$ goes to zero. In all these setups, in the limit of electrically small transverse dimensions, the product $\bar{\beta}\bar{d}$ remains constant. Further analogies among these setups will be discussed in future works.

## *3.    Realistic models for the particles forming the chain*

We have shown in the previous section how the nanoparticles composing the linear chain under analysis are well described by their effective electric polarizability in the regime of interest here. $\alpha_{ee}$ is strictly related to the geometry of the particles and to the permittivity of their material(s). Consider, for simplicity, the case of a chain of spherical homogeneous particles of radius $a$ and permittivity $\varepsilon$. In general their polarizability may be obtained from (1) and the formulas given in [29] for $c_1^{TM}$, but in the quasi-static limit in which the spheres are much smaller than the wavelength, as in the case of interest here, one can write in the lossless case [1]-[3]:

$$\text{Re}\left[\bar{\alpha}_{ee}^{-1}\right] = \frac{3}{2}(k_0 a)^{-3} \frac{\varepsilon + 2\varepsilon_0}{\varepsilon - \varepsilon_0}. \qquad (23)$$

Notice that, when the particles are too small, the permittivity of the bulk material may not be adequate to describe the quantum effects associated with the nanoparticles, but a corrected value of $\varepsilon$, distinct form that of the bulk material (usually with a higher attenuation factor), may be employed [47]. We do not enter further into these discussions here, since it is out of the scope of the present paper, but in such a case we note that Eq. (23) may still take into account these considerations.

*a) Varying the permittivity of the particles*

It is interesting to see that due to the geometrical requirement $d > 2a$ and to the fact that $\dfrac{\varepsilon + 2\varepsilon_0}{\varepsilon - \varepsilon_0} > 1$ for any $\varepsilon > \varepsilon_0$, a chain made of conventional dielectric particles in empty space cannot guide any concentrated guided beam around the chain, since $\left(\bar{d}^3 \operatorname{Re}\left[\bar{\alpha}_{ee}^{-1}\right]\right) > 12$ for any $\varepsilon > \varepsilon_0$ (see Fig. 2 and 4). The region of guidance is expected to be around the resonance of the individual particles, which is at around $\varepsilon = -2\varepsilon_0$.

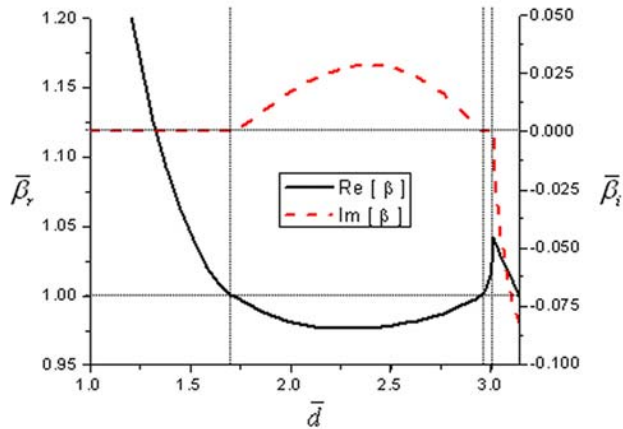

Figure 8 – Dispersion of the wave number for the longitudinal polarization in terms of the particle spacing in a chain of spherical particles with $\varepsilon = -2\varepsilon_0$, i.e., at their individual resonance.

Figure 8, as an example, shows the variation of the real and imaginary parts of $\bar{\beta}$ for the longitudinal polarization, as a function of the spacing $\bar{d}$ for a chain of particles at their individual resonance, i.e., with $\mathrm{Re}\left[\bar{\alpha}_{ee}^{-1}\right]=0$, which in this case happens around $\varepsilon=-2\varepsilon_0$. A similar plot may be drawn for the transverse polarization. For small spacing, not reported in the figure, the guided mode is more and more confined, and the hyperbolic behavior of the curve is well approximated by (21), i.e., $\bar{\beta}=\pi/(2\bar{d})$. However, when $\bar{d}=1.71$, as noted in the previous section, the mode starts leaking out energy, since Eq. (13) is not satisfied any more. In this leaky-wave region $\bar{\beta}$ has an imaginary part, which is positive since the mode in this case is a forward improper leaky mode. Again, consistently with Fig. 2 and the previous discussion, there is another small region of spacing where a mode can be guided, with a value of $\bar{\beta}$ increasing with increasing $\bar{d}$. When $\bar{\beta}=\pi/\bar{d}$ the mode enters a cut-off region where the mode does not propagate and the phases of neighboring dipoles remain 180° out of phase with each other, with the expression for $\mathrm{Re}\left[\bar{\beta}\right]=\pi/\bar{d}$. Again an attenuation factor is present in this region, since no propagation is admitted and the chain is in its stop-band. The sign of the imaginary part in this case is irrelevant, since complex conjugates solutions are admitted in this stop-band region.

In this sub-wavelength limit, the regions of guidance in terms of the particle permittivity may be easily written as:

$$\frac{f_1/\eta+3}{f_1/\eta-3/2}<\varepsilon<\frac{f_2/\eta+3}{f_2/\eta-3/2} \qquad (24)$$

where $f_1 = 3\left[\xi(3) + Cl_3(2\bar{d}) + \bar{d}\, Cl_2(2\bar{d})\right]$, $f_2 = 6\left[Cl_3(\bar{d}+\pi) + \bar{d}\, Cl_2(\bar{d}+\pi)\right]$ for the longitudinal case and $f_1 = -3\left[Cl_3(\bar{d}+\pi) + \bar{d}\, Cl_2(\bar{d}+\pi) - \bar{d}^2 Cl_1(\bar{d}+\pi)\right]$, $f_2 = \bar{d}^3 \bar{\alpha}_{min}^{-1}$ for the transverse polarization. Moreover, $\eta = d/a$ represents the ratio between the particle radius and their center-to-center distance (therefore $\eta > 2$ for geometrical reasons).

As an example, Fig. 9 plots this range of guidance for the two polarizations and for different values of $\eta$.

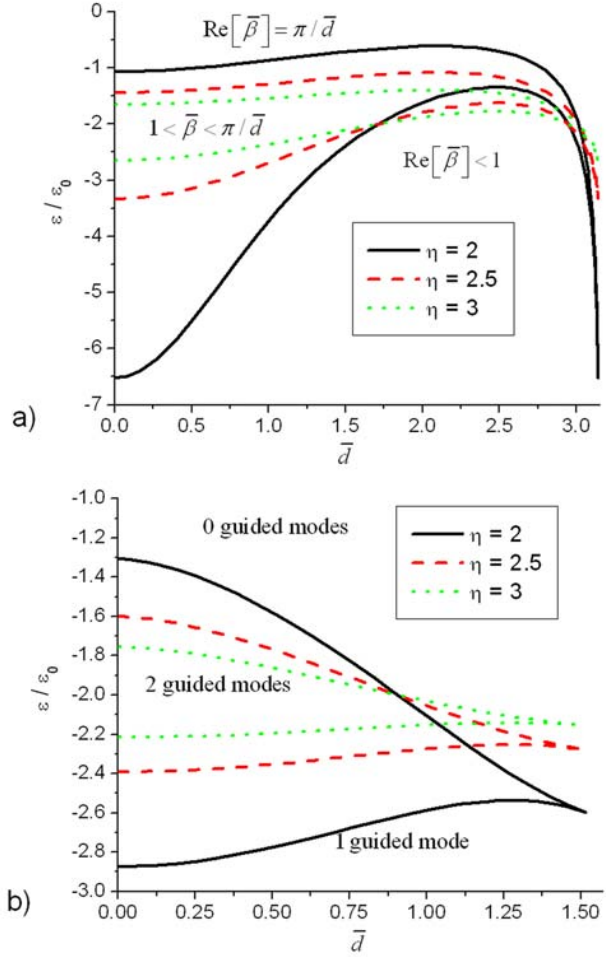

Figure 9 – Range of the required permittivity of the spheres and their spacing for having guided modes for (a) the longitudinal polarization, and (b) the transverse polarization.

Fig. 9a shows the range of required permittivities for having a guided mode with longitudinal polarization, whereas Fig. 9b shows the region in which the confined backward-wave guided mode with transverse polarization is supported by the chain (which coincides with the two-mode region). The figures show some interesting features. First, even though the $Q$ of the dipolar resonance of a plasmonic particle increases dramatically when its size decreases [29] consistent with the Chu limit [37], and therefore its resonance rapidly becomes very concentrated in a narrow range of permittivities near $\varepsilon = -2\varepsilon_0$ in the sub-wavelength limit, the closely-spaced chains of small particles do not necessarily require for their particles to have permittivity so close to this value, since the range of required polarizability diverges as $\bar{d}^{-3}$ for small spacing. These two factors interestingly compensate each other, and therefore for very small closely-spaced particles the range of necessary permittivities for guidance remains a finite and relatively large range for both polarizations. This is due to the strong coupling of the individual resonances in the chain, which allows broadening of the individual bandwidth, similar to what happens in a transmission line metamaterial [48]-[49]. We are currently working on a detailed analysis of this coupling issue, which will be presented elsewhere and that may be exploited for different applications.

When the spacing is increased, the range of possible permittivities to yield a resonant behavior narrows down around the value of $\varepsilon = -2\varepsilon_0$, since the coupling between the particles is diminished and the array resonance is dominated by the individual resonances of the particles. The limiting values may be interestingly calculated in closed form in the limit of $\bar{d} \to 0$ and are given as follows:

$$L: 1 - \frac{3\eta}{\eta - 4\xi(3)} < \frac{\varepsilon}{\varepsilon_0} < 1 - \frac{3\eta}{\eta + 3\xi(3)}$$
$$T: 1 + \frac{6\eta}{-2\eta + 3\xi(3)} < \frac{\varepsilon}{\varepsilon_0} < 2 - \frac{4\eta}{\eta + \xi(3)} \quad , \quad (25)$$

whereas at the upper limit, which is $\bar{d} \to \pi$ for the longitudinal polarization and $\bar{d} \to \hat{d}$ for the transverse case, the only possible value for the resonance remains:

$$L: \frac{\varepsilon}{\varepsilon_0} = 1 - \frac{3\eta}{\eta - 4\xi(3)}$$
$$T: \frac{\varepsilon}{\varepsilon_0} = -2 + \frac{3.982}{-\eta + 1.327} \quad , \quad (26)$$

which is the value of permittivity that leads to a resonance in an array of spherical particles with spacing of $\lambda_0/2$, in both polarizations. One may notice also how all the curves in the case of longitudinal polarization meet at the point where $\varepsilon = -2\varepsilon_0$ and $\bar{d} = 1.71$, since for this value of spacing the longitudinal mode has its cut-off at the resonance of the particles, which, in the quasi-static limit where (23) is valid, is independent of the size of the particles. The transverse mode has a similar property at $\bar{d} \simeq \hat{d}/2$.

By increasing the value of $\eta$, all quantities in (25) and (26) tend to $\varepsilon = -2\varepsilon_0$, since when we increase the spacing or decrease the size of the particles we expect to narrow down the bandwidth of the guided mode, thus concentrating on the region of guidance around the frequencies for which $\text{Re}\left[\bar{\alpha}_{ee}^{-1}\right] \simeq 0$. The case of the longitudinal polarization offers better performance than the transverse polarization in terms of the range of permittivities for supporting a guided mode. This is analogous to the other metamaterial setups for wave guiding, (e.g., [41]-[44], [48]-[49]), in which the operation in the backward regime usually has a relatively smaller bandwidth.

*b) Dispersive materials*

We have shown in the previous paragraph how a confined guided mode that is traveling along a linear chain of particles may be obtained under the condition that the particles in the chain are plasmonic, i.e., with negative permittivity. As is well known, however, presence of negative values of the constitutive parameters necessarily implies a non-negligible dispersion with frequency and thus the presence of material absorption [1]. Here therefore we consider certain realistic models for the frequency dispersion of the materials and we verify and augment the predictions of the previous section under these assumptions.

As an example, let us consider a chain of spherical particles with Drude-model permittivity, i.e., $\varepsilon(\omega) = \varepsilon_0 \left(1 - \dfrac{3\omega_0^2}{\omega^2}\right)$ for their materials, surrounded by free space. In Fig. 10 the dispersion of the guided $\beta$ versus frequency is plotted for the two cases of polarizations for particles with radius $a = \lambda_0 / 20$ (calculated at $\omega = \omega_0$) and spacing $d = 2.5a$. The horizontal axis in the figure indicates the normalized frequency (i.e., the frequency normalized with respect to the central frequency $\omega = \omega_0$, at which the particles would experience their individual resonance, since $\varepsilon = -2\varepsilon_0$). The two curves are limited in the maximum value by the line with $\beta = \pi/d$ and in the minimum value by the light line $\beta = k_0$. The locus $\beta = \beta_{min}$ is also reported in the figure, which meets the red dashed line (transverse polarization) at the point where the two dispersion curves meet. This point can be considered the upper limit of the guidance regime in the case of transverse polarization. Notice that the vertical axis is normalized to the wave number in free space at the central frequency $\omega_0$.

The slope of the two curves clearly confirms the conclusions of the previous sections regarding the behavior of the group velocity in the two polarizations. One notices how the backward-wave mode (transverse polarization) has a smaller bandwidth (about 10% of the central frequency) when compared with the bandwidth of the forward-wave mode (almost 30%). The bandwidth can be further increased by reducing $d$. However, consistent with the previous analysis and with Fig. 9, a change in the particle size while keeping $\eta$ fixed does not appreciably affect this aspect as long as losses are negligible.

It should be mentioned how the values of bandwidth obtained here for the backward, as well as forward, regimes are larger than those usually achievable in a left-handed metamaterial made of resonant inclusions with an effective negative refraction. This provides an interesting potential for extending these concepts to 2-D or 3-D collections of such closely-packed plasmonic particles, which, when properly combined and under proper excitation, may constitute an alternative way of building broader bandwidth left-handed metamaterials with the desired effective refraction at microwave, infrared and optical frequencies [50]. The concepts presented here are in many aspects the 1-D analogue of the planar geometry reported in [41], i.e., a 1-D nanotransmission line with features analogous to broad-band negative-index transmission-line metamaterials at microwave frequencies [48]-[49], but extendable to infrared and optical frequencies. We discuss this point in the next paragraph.

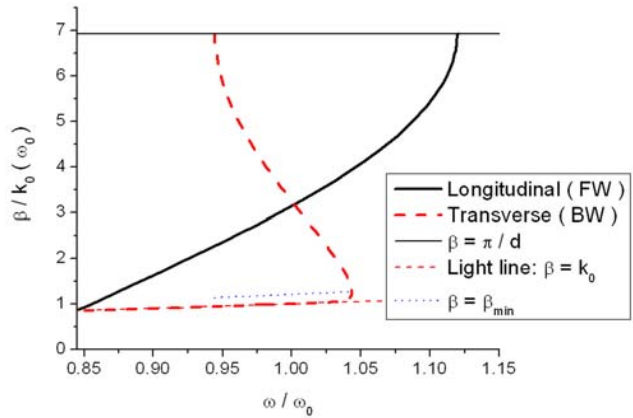

Figure 10 - Frequency dispersion for a chain of spherical particles with radius $a = \lambda_0 / 20$ (calculated at $\omega = \omega_0$), spacing $d = 2.5 a$ and permittivity following the Drude model $\varepsilon(\omega) = \varepsilon_0 \left( 1 - 3\omega_0^2 / \omega^2 \right)$. The guided wave number $\beta$ is normalized in the vertical axis to the wave number in free-space $k_0$ calculated at $\omega = \omega_0$.

*c) Nanocircuit interpretation*

The interpretation of this chain configuration in terms of nanotransmission line is more than a mere analogy, as highlighted above. Indeed, this concept can be fully motivated in terms of our nanocircuit interpretation of the light interaction with plasmonic particles [51]. It is well known that at low frequencies properly arranged circuit elements, namely, the cascades of inductors and capacitors, may constitute transmission lines that guide energy without (or with small) radiation losses. At the IR and optical frequencies, when conducting materials exhibit different material properties, the displacement current may take the role of conduction current, and the nanocircuit elements have to be re-envisioned based on a different physical mechanism. In particular, we have shown in [51] how plasmonic and non-plasmonic nanoparticles may be envisioned as nanoinductors and nonplasmonic elements, respectively. In the case of interest here, the chains of plasmonic particles interleaved and surrounded with non-plasmonic gaps may therefore be interpreted as the cascade of such nanoinductors and nanocapacitors that constitute nano-transmission lines in the optical regime when suitably designed and properly excited. In the longitudinal propagation, due to the polarization of the electric field, the plasmonic particles may be heuristically regarded as the "series" nanoinductors surrounded by "parallel" nano-capacitors, providing an equivalent L-C transmission line with forward-wave behavior, whereas in the transverse polarization the roles of inductors and capacitors are essentially interchanged, providing a 1-D backward-wave transmission line, analogous to those in the microwave negative-index TL metamaterials. This is consistent with the findings of the previous section. An extension to 2-D and 3-D may therefore allow synthesizing effective 2-D and 3-D nanotransmission line negative-index metamaterials at optical frequencies [50], analogous to those synthesized at microwaves using lumped circuit elements [48]-[49]. The advantages of broad bandwidth and robustness to losses, which are typical of such metamaterials at microwave frequencies, are expected to also apply to such optical nano-metamaterials, as these 1-D results confirm and our preliminary results on the 3-D geometry show [50]. We will present a complete analysis of such bulk optical nanotransmission line metamaterials in the near future.

It is worth noting that we have also utilized similar concepts in order to envision and explain the sub-diffraction low-attenuation propagation along thin planar layers or cylindrical rods of plasmonic and non-plasmonic materials [41]-[42]. These cases may be seen as the limit of vanishing gaps in the nanoparticle chains presented here. An extended analysis of the analogies of these array problems with the transmission-line circuit theory will be presented in a future work.

*c) Material losses*

The behavior of these structures in the presence of material losses has been already analyzed in the previous section, and the dependence of the attenuation factor $\bar{\beta}_i$ on losses has been found to be directly proportional to the quantity $\bar{\alpha}_{loss}^{-1}$. Here we embed the presence of losses in the material permittivity. This allows evaluating how the size of the particles that form the chain may play an important role as a lower limit on squeezing and confining the guided beam below the diffraction limit.

Adding to (23) the contribution of material losses, we can write for a single particle with complex permittivity $\varepsilon = \varepsilon_r + i\varepsilon_i$:

$$\bar{\alpha}_{loss}^{-1} = \frac{9\varepsilon_i}{2} \frac{\varepsilon_0 (k_0 a)^{-3}}{(\varepsilon - \varepsilon_0)^2 + \varepsilon_i^2}. \qquad (27)$$

If $\bar{\alpha}_{loss}^{-1}$ grows with increasing the imaginary part of the permittivity, as physically expected, it also increases with the inverse volume of the particle. Therefore, if reducing the size of the particles while keeping $\eta$ fixed may increase the bandwidth of guided propagation along the chain (since the spacing between them should be accordingly reduced to allow a fixed $\eta$), a lower limit is represented by (27) combined with (22), which ensures that there is a physical lower limit on squeezing the beam in a sub-wavelength scale with the presence of realistic losses. Therefore, as physically expected, there is a trade-off between bandwidth and sub-diffraction on the one hand, and the sensitivity to losses on the other hand. This is also physically justifiable due to the concentration of the guided beam in a sub-wavelength region where material losses are present.

Eq. (22) becomes:

$$L: \bar{\beta}_i = \frac{3\varepsilon_i}{4} \frac{\varepsilon_0 (k_0 a)^{-1} \eta^2}{(\varepsilon - \varepsilon_0)^2 + \varepsilon_i^2}$$

$$T: \bar{\beta}_i = -\frac{3\varepsilon_i}{2} \frac{\varepsilon_0 (k_0 a)^{-1} \eta^2}{(\varepsilon - \varepsilon_0)^2 + \varepsilon_i^2}, \quad (28)$$

which shows how the attenuation factor in this configuration is inversely proportional to the linear size of the particles, much better than the volume (cubic) dependence that an isolated plasmonic resonance generally shows. Clearly, for a given size of the particles, more robustness to losses is ensured when the particles are closely packed, as the presence of $\eta^2$ shows in the numerator of (28).

*d) Arbitrary polarization*

Elliptically polarized modes propagating along the chain may be obtained by exciting both longitudinal and transverse components of the polarizable particles. Assuming an isotropic polarizability, and referring to Fig. 1, the polarization of the exciting dipole at $x = 0$ may be written, without lose of generality, as:

$$\mathbf{p}_0 = a_l \hat{\mathbf{x}} + a_t \hat{\mathbf{z}} \qquad (29)$$

where $a_l$ and $a_t$ are, respectively, the complex amplitudes driving the longitudinal and the transverse polarization. Owing to the linearity of the problem, the two modal polarizations of the system may be analyzed independently in this case. Along the same direction, e.g., the positive $x$, we would experience the propagation of two orthogonal modes, the longitudinal one with positive phase and group velocity, i.e., $\beta_l > 0$ and the transverse one with $\beta_t < 0$ (here we neglect the possible excitation of the spurious non-confined mode in the transverse polarization). The generic amplitude of the $N$-th dipole would become:

$$\mathbf{p}_N = a_l e^{iN\beta_l d} \hat{\mathbf{x}} + a_t e^{-iN|\beta_t|d} \hat{\mathbf{z}}, \qquad (30)$$

which in general is arbitrarily polarized. Interesting combinations may be envisioned. For instance, a combination for which $a_l = a_t$ and $\beta_l = -\beta_t$, can be designed as the chain of Fig. 10 at the frequency where the two curves meet with $\mathbf{p}_N = a_l \left[ \cos(N\beta_l d)(\hat{\mathbf{x}} + \hat{\mathbf{z}}) + i \sin(N\beta_l d)(\hat{\mathbf{x}} - \hat{\mathbf{z}}) \right]$. This would have the interesting property of being linearly polarized at 45° from the axis when $d$ is a multiple of $\pi/(2\beta_l)$, rotating its orientation every particle, or circularly polarized for $\beta_l d = \pi/4 + L\pi/2$ with $L$ being an integer number. Also a circularly polarized field at $\mathbf{p}_0$, i.e., $a_l = \pm i\, a_t$ would produce a rotating circular polarization for $\beta_l d = L\pi/2$ or a rotating linear polarization for $\beta_l d = \pi/4 + L\pi/2$.

*d) Realistic plasmonic materials*

From our previous discussions above, we have seen that chains of metamaterial or noble metal particles with negative permittivity may provide interesting potentials for guiding a sub-diffraction beam with a relatively low attenuation factor (we reiterate here that μ-negative (MNG) materials may be employed as well, since by duality they would support coupled magnetic resonances). However, in analyzing the possibility of applying these concepts to a real-life setup, we have to deal with the required properties of the materials of the particles. As previously shown, the chain of dielectric particles made of standard materials would not provide the required guiding properties. Metamaterials may be properly synthesized by embedding resonant inclusions in a host dielectric in order to have the desired effective negative properties [52]-[55]. Clearly, the size of such inclusions should be much smaller than that of the metamaterial particle in order to let the particle be considered as a bulk material with negative parameters. Therefore, in dealing with sub-wavelength particles, the inclusions that "make" the materials may be required to be much smaller than the operating wavelength, and still produce a sizeable resonance. This may represent a challenging task for the designer. Even the anisotropy of the metamaterials may represent an issue in this sense, since the inclusions may have preferred directions for their interaction with the impinging field. Homogeneous spherical particles, however, are polarized with a uniform field inside their overall volume at the dipolar resonance, and therefore it would be sufficient to orient the axis of anisotropy in the direction of the expected polarization of the field in order to obtain results consistent with this analysis.

If metamaterials may provide a flexible way of tailoring the material properties at the desired frequency of operation, at infrared and optical frequencies noble metals and polar dielectrics naturally possess the required isotropic negative permittivity [1]-[6]. On the one hand this is useful in envisioning a direct and easy application of the present analysis, but still the set of noble metals is limited in terms of the possible material dispersion and the range of frequencies at which the desired values of permittivity are met with possibly low losses. Additional degrees of freedom in the geometry of the particles of the chain may provide possibilities for tuning the properties of these guided modes at the desired frequency. For instance, following [29], core-shell concentric spherical nanoparticles may suggest a way for tailoring the properties of the guided modes along the chain. We have found that the choice of the proper filling ratio (i.e., the ratio of radii) of core-shell spheres partially filled with ENG or MNG materials may allow tuning the particle electric or magnetic polarizability in frequency.

Following (1), one can write $\text{Re}\left[\bar{\alpha}_{ee}^{-1}\right] = \text{Im}\left[\left(c_1^{TM}\right)^{-1}\right]$, which in the case of lossless particles is simplified into $\text{Re}\left[\bar{\alpha}_{ee}^{-1}\right] = V_1^{TM}/U_1^{TM}$. Using the closed form expressions for these quantities derived in [29], one can evaluate the exact dispersion for a chain of core-shell spherical particles.

Let us consider for instance the case of a chain of particles of radius $a = 15\,nm$ with spacing $d = 75\,nm$ excited at optical wavelength of $\lambda_0 = 500\,nm$. Assuming homogeneous spherical particles in the chain, the modal dispersion versus the permittivity is reported in Fig. 11a for the case of longitudinal polarization, where one can note the range of permittivities over which the chain would support a guided mode. A too low permittivity (in this case below $\varepsilon = -2.16\,\varepsilon_0$ would start leaking energy out of the chain, and a too high permittivity, i.e., for $\varepsilon > -1.97\,\varepsilon_0$, would enter the cut-off region of the array, where $\bar{\beta}_r = \pi/\bar{d}$ and the attenuation factor increases with $\varepsilon$. Clearly, the range of frequencies for which a realistic material may be available with these required values of permittivities might be narrow and may not coincide with the frequency of interest (not to mention the fact that this range may not necessarily correspond to a low-loss range of frequencies). Imagine we are willing to use spherical particles made of silver, which at $\lambda_0 = 500\,nm$ have a real part of permittivity given by $\mathrm{Re}\left[\varepsilon_{Ag}\right] = -9.77\,\varepsilon_0$ [56]. Silver has a relatively low material absorption at this frequency, so it may appear to be a suitable material for these purposes. However, the real part of permittivity clearly does not fall in the range of allowable permittivities, as Fig. 11a shows. However, if we cover silicon carbide spheres ($\varepsilon_{SiC} = 6.52\,\varepsilon_0$ [57]) with silver shells, maintaining the same outer radius for the overall particle, we can move the resonance of the core-shell particles to the desired frequency, as found in [29]. Fig. 11b shows the variation of $\beta$ with the ratio of radii $\gamma = a_1/a$ at the frequency of interest, i.e., at $\lambda_0 = 500\,nm$.

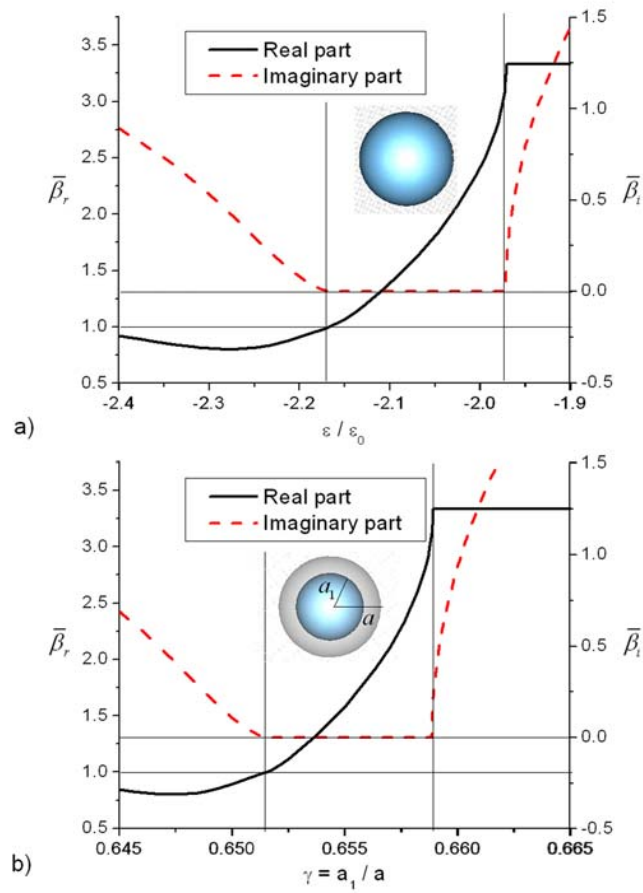

Figure 11 - (a) Variation of normalized $\bar{\beta}_r$ in terms of the material permittivity for a chain of homogeneous spherical particles with radius $a = 15\,nm$ and spacing $d = 5a$ at the wavelength $\lambda_0 = 500\,nm$ for the longitudinal polarization; (b) Variation of normalized $\bar{\beta}_r$ in terms of the ratio of radii $\gamma$, for the same size, spacing and polarization, but for a $SiC$ sphere covered with $Ag$ shell.

In this example, the spacing among the particles is pretty large, and therefore still the sensitivity to the material parameters of the chain is relatively high in order to get a low-attenuation mode, but this shows how in principle it is possible to add additional degrees of freedom to the problem in order to adjust the frequency of operation and its properties. Here the degree of freedom is provided by the filling ratio $\gamma$, but in other cases it may be represented by the shape of the object and its eccentricity (for instance for ellipsoidal particles). It is clear how by properly choosing the materials in the regions where they show sufficiently low losses, together with the proper choice of the chain geometrical parameters in the regions with minimal loss influence (20), it may in principle be possible to tailor modes that can propagate for relatively long distances in a sub-diffractive mode. As a final example, Fig. 12 shows the real and imaginary parts of guided wave numbers for longitudinal modes supported by a chain of spherical particles made of silver, using the realistic experimental data available in the literature [56] for bulk silver at optical frequencies with $a = 10\,nm$, $d = 22.5\,nm$, as a function of the wavelength of operation $\lambda_0$. Operating around $\lambda_0 = 370\,nm$ can lead to a guided wave propagation with a reasonable good ratio of $\bar{\beta}_r / \bar{\beta}_i$, providing the possibility of sub-diffraction propagation with a relatively low attenuation.

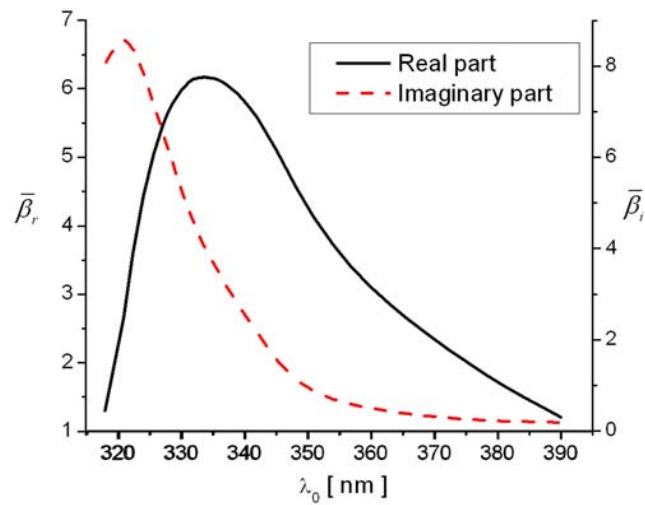

Fig. 12 – Frequency dispersion of the longitudinally-polarized modes supported by a chain of spherical silver particles with $a = 10\,nm$ and $d = 22.5\,nm$, considering realistic experimental data available in the literature for the bulk silver [56].

## 4. Conclusions

In this paper we have analyzed in detail the various aspects of guiding properties of linear arrays of metamaterial/plasmonic particles. After deriving novel closed-form analytical dispersion relations for the sub-diffractive guided modes of such arrays, we have been able to provide conditions for minimal absorption and maximum bandwidth of such modes. A discussion on the realistic possibility of realizing such setups, with numerical examples considering ohmic absorption and frequency dispersion has also been provided. This may open new doors in the realization of microwave, infrared and optical nanotransmission lines and nano-waveguides. Enlightened by the nanocircuit interpretation of this phenomenon, also an extension to 2-D and 3-D setups for the realization of broader band left-handed metamaterials at high frequencies has been envisioned and forecasted.

**Acknowledgements**

This work is supported in part by the U.S. Air Force Office of Scientific Research (AFOSR) grant number FA9550-05-1-0442. Andrea Alù was partially supported by the 2004 SUMMA Graduate Fellowship in Advanced Electromagnetics.

**REFERENCES**

[1] L. Landau, and E. M. Lifschitz, *Electrodynamics of Continuous Media* (Pergamon Press, Oxford, UK, 1984).

[2] C. F. Bohren, and D. R. Huffman, *Absorption and Scattering of Light by Small Particles* (Wiley, New York, 1983).

[3] J. D. Jackson, *Classical Electrodynamics* (Wiley, New York, USA, 1975).

[4] J. Gómez Rivas, C. Janke, P. Bolivar, and H. Kurz, *Opt. Express*, vol. 13, pp. 847 (2005).

[5] G. D. Mahan, G. Obermair, "Polaritons at surfaces," *Physical Review*, vol. 183, no. 3, pp. 834-841, July 15, 1969.

[6] M. Kerker, "Founding fathers of light scattering and surface-enhanced Raman scattering," *Applied Optics*, vol. 30, no. 33, pp. 4699-4705, November 1991.

[7] M. Quinten, A. Leitner, J. R. Krenn, F. R. Aussenegg, "Electromagnetic energy transport via linear chains of silver nanoparticles", *Optics Letters*, vol. 23, no. 17, 1331 (1998).

[8] S. A. Tretyakov, and A. J. Vitanen, "Line of periodically arranged passive dipole scatterers," *Electrical Engineering,* vol. 82, pp. 353-361, 2000.

[9] M. L. Brongersma, J. W. Hartman, and H. A. Atwater, "Electromagnetic energy transfer and switching in nanoparticle chain arrays below the diffraction limit," *Physical Review B*, vol. 62, no. 24, pp. 16356-16359, Dec. 2000.

[10] S. A. Maier, M. L. Brongersma, and H. A. Atwater, "Electromagnetic energy transport along arrays of closely spaced metal rods as an analogue

to plasmonic devices," *Applied Physics Letters*, vol. 78, no. 1, pp. 16-18, Jan. 2001.

[11] J. R. Krenn, M. Salerno, N. Felidj, B. Lamprecht, G. Schider, A. Leitner, F. R. Aussenegg, J. C. Webber, A. Dereux, and J. P. Goudonnet, "Light field propagation by metal micro- and nanostructures," *Journal of Microscopy*, vol. 202, pp. 122-128, April 2001.

[12] S. A. Maier, P. G. Kik, and H. A. Atwater, "Observation of coupled plasmon-polariton modes of plasmon waveguides for electromagnetic energy transport below the diffraction limit," *Proc. SPIE, Properties of Metal Nanostructures*, N. J. Halas, ed., vol. 4810, pp. 71-81 (2002).

[13] A. D. Yaghjian, "Scattering-matrix analysis of linear periodic arrays," *IEEE Transactions on Antennas and Propagation*, vol. 50, no. 8, pp. 1050-1064, Aug. 2002.

[14] S. A. Maier, P. G. Kik, and H. A. Atwater, "Observation of coupled plasmon-polariton modes in Au nanoparticle chain waveguides of different lengths: estimation of waveguide loss," *Applied Physics Letters*, vol. 81, no. 9, pp. 1714-1716, Aug. 2002.

[15] S. A. Maier, P. G. Kik, and H. A. Atwater, "Optical pulse propagation in metal nanoparticle chain waveguides," *Physical Review B*, vol. 67, 205402, May 6, 2003.

[16] S. A. Maier, P. G. Kik, H. A. Atwater, S. Meltzer, E. Harel, B. E. Koel, and A. A. G. Requicha, "Local detection of electromagnetic energy

transport below the diffraction limit in metal nanoparticle plasmon waveguides," *Nature Materials*, vol. 2, pp. 229-232, April 2003.

[17] S. K. Gray, and T. Kupka, "Propagation of light in metallic nanowire arrays: finite-difference time-domain studies of silver cylinders," *Physical Review B*, vol. 68, 045415 (2003).

[18] R. Arias, and D. L. Mills, "Collective modes of interacting dielectric spheres," *Physical Review B*, vol. 68, 245420 (2003).

[19] S. Y. Park, and D. Stroud, "Surface-plasmon dispersion relations in chains of metallic nanoparticles: an exact quasi-static calculation," *Physical Review B*, vol. 69, 125418 (2004).

[20] W. H. Weber, and G. W. Ford, "Propagation of optical excitations by dipolar interactions in metal nanoparticle chains," *Physical Review B*, vol. 70, 125429 (2004).

[21] V. A. Markel, "Divergence of dipole sums and the nature of non-Lorentzian exponentially narrow resonances in one-dimensional periodic arrays of nanospheres," *J. Phys. B: At. Mol. Opt. Phys.*, vol. 38, L115-L121 (2005).

[22] R. A. Shore, and A. D. Yaghjian, "Traveling electromagnetic waves on linear periodic arrays of small lossless penetrable spheres," *Proc. of 2004 International Symposium on Antennas and Propagation (ISAP'04)*, Sendai, Japan, pp. 425-428.

[23] R. A. Shore, and A. D. Yaghjian, "Travelling electromagnetic waves on linear periodic arrays of lossless spheres," *Electronics Letters*, vol. 41, no. 10, May 2005.

[24] R. A. Shore, and A. D. Yaghjian, "Traveling electromagnetic waves on linear periodic arrays of lossless penetrable spheres," *IEICE Trans. Commun.*, vol. E88-B, no. 6, pp. 2346-2352, June 2005.

[25] D. S. Citrin, "Plasmon-polariton transport in metal-nanoparticle chains embedded in a gain medium," *Optics Letters*, vol. 31, no. 1, pp. 98-100, Jan. 1, 2006.

[26] A. F. Koenderink, and A. Polman, "Complex response and polariton-like dispersion splitting in periodic metal nanoparticle chains," *Physical Review B*, vol. 74, 033402 (2006).

[27] A. Alù, and N. Engheta "Polarizabilities and effective parameters for collection of spherical nano-particles containing concentric double-negative or single-negative shells," in *Proceedings of the 2004 URSI International Symposium on Electromagnetic Theory*, Pisa, Italy, pp. 24-26, May 23-27, 2004.

[28] A. Alù, and N. Engheta, "Low-damping guided modes along nano-transmission lines with chains of quadrupolar resonant plasmonic nano-particles," in *Proceedings of the 28th General Assembly of the International Union of Radio Science (URSI)*, New Delhi, India, Paper No. 99, October 23-29, 2005.

[29] A. Alù, and N. Engheta, "Polarizabilities and effective parameters for collections of spherical nano-particles formed by pairs of concentric double-negative (DNG), single-negative (SNG) and/or double-positive (DPS) metamaterial layers," *Journal of Applied Physics*, vol. 97, 094310 (12 pages), May 1, 2005.

[30] L. Lewin, *Polylogarithms and Associated Functions*, New York, Elsevier North-Holland, 1981.

[31] Wolfram Mathematica$^{TM}$ 5.2, http://www.wolfram.com

[32] R. E. Collin, and F. J. Zucker, *Antenna Theory* (McGraw-Hill ed., 1969).

[33] P. A. Balov, and C. Simovski, "Homogenization of electromagnetic crystals formed by uniaxial resonant scatterers," *Physical Review E*, vol. 72, 026615, August 2005.

[34] I. A. Stegun, "Miscellaneous functions," in M. Abramowitz, I. A. Stegun, *Handbook of Mathematical Functions* (Dover Publications, Inc., New York, 1970).

[35] L. B. Felsen, and N. Marcuvitz, *Radiation and Scattering of Waves* (IEEE Press, 1994).

[36] M. Bertolotti, "Wave interactions in photonic band structures: an overview," *Journal of Optics A: Pure and Applied Optics*, vol. 8, pp. S9-S32 (2006).

[37] L. J. Chu, "Physical limitations of omni-directional antennas," *J. Appl. Phys.* 19, 1163, 1948.

[38] T. Tamir, and A. A. Oliner, "The spectrum of electromagnetic waves guided by a plasma layer," *IRE Transactions on Antennas and Propagation*, vol. 11, pp. 317-332, Feb. 1963.

[39] S. R. Laxpati, and R. Mittra, "Energy considerations in open and closed waveguides," *IEEE Transactions on Antennas and Propagation*, vol. 13, no. 6, pp. 883-890, Nov. 1965.

[40] P. A. Belov, R. Marques, S. I. Maslovski, I. S. Nefedov, M. Silveirinha, C. R. Simovski, and S. A. Tretyakov, "Strong spatial dispersion in wire media in the very large wavelength limit," *Physical Review B*, vol. 67, 113103 (2003).

[41] A. Alù, and N. Engheta, "Optical nano-transmission lines: synthesis of planar left-handed metamaterials in the infrared and visible regimes," *Journal of the Optical Society of America B, Special Focus Issue on Metamaterials*, Vol. 23, no. 3, pp. 571-583, March 2006.

[42] A. Alù, and N. Engheta "Anomalies in the surface wave propagation along double-negative and single-negative cylindrical shells," a talk presented at *the Progress in Electromagnetics Research Symposium (PIERS'04)*, Pisa, Italy, one-page abstract in the CD Digest, March 28-31, 2004.

[43] A. Alù, and N. Engheta, "An overview of salient properties of planar guided-wave structures with double-negative (DNG) and single-negative (SNG) layers," in *Negative Refraction Metamaterials: Fundamental Properties and Applications*, G. V. Eleftheriades, and K. G. Balmain, eds.,

IEEE Press, John Wiley & Sons Inc., Hoboken, New Jersey, pp. 339-380, 2005.

[44] N. Engheta, A. Alù, R. W. Ziolkowski, A. Erentok, "Fundamentals of waveguide and antenna applications involving DNG and SNG metamaterials," in *Metamaterials: Physics and Engineering Explorations*, N. Engheta and R. Ziolkowski, eds., IEEE Press, John Wiley and Sons, Inc., pp. 43-86, May 2006.

[45] C. R. Simovski, A. J. Vitanen, and S. A. Tretyakov, "Resonator mode in chains of Silver spheres and its possible application," *Physical Review E*, vol. 72, 066606 (2005).

[46] C. R. Simovski, A. J. Vitanen, and S. A. Tretyakov, "A resonator mode in linear arrays of silver spheres and cylinders," *J. Zhejiang Univ. Science A*, vol. 7, no. 1, pp. 29-33 (2006).

[47] P. Mulvaney, "Surface plasmon spectroscopy of nanosized metal particles," *Langmuir*, vol. 12, pp. 788-800 (1996).

[48] G. V. Eleftheriades, A. K. Iyer, and P. C. Kremer, "Planar negative refractive index media using periodically L–C loaded transmission lines," *IEEE Transactions on Microwave Theory and Techniques*, vol. 50, pp. 2702-2712 (2002).

[49] L. Liu, C. Caloz, C.-C. Chang, and T. Itoh, "Forward coupling phenomena between artificial left-handed transmission lines," *Journal of Applied Physics*, vol. 92, pp. 5560-5565 (2002).

[50] A. Alù, and N. Engheta, "Dispersion properties of volumetric optical nanotransmission-line metamaterials with negative refraction," a talk presented at the *USNC/CNC/URSI National Radio Science Meeting*, Albuquerque, NM, USA, one-page abstract in p. 542 of the digest of URSI abstracts, July 9-14, 2006.

[51] N. Engheta, A. Salandrino, and A. Alù, "Circuit elements at optical frequencies: nano-inductors, nano-capacitors and nano-resistors," *Physical Review Letters*, Vol. 95, 095504, August 26, 2005.

[52] W. Rotman, "Plasma simulation by artificial dielectrics and parallel-plate media," *IRE Trans. Antennas Propagat.*, vol. 10, pp. 82-84, 1962.

[53] J. B. Pendry, A. J. Holden, D. J. Robbins, and W. J. Stewart, "Low-frequency plasmons in thin wire structures," *J. of Physics: Condensed Matter*, vol. 10, pp. 4785-4809, 1998.

[54] J. B. Pendry, A. J. Holden, D. J. Robbins, and W. J. Stewart, "Magnetism from conductors and enhanced nonlinear phenomena," IEEE *Trans. Microwave and Theory Techniques*, vol. 47, no. 11, pp. 2075-2081, Nov. 1999.

[55] P. Gay-Balmaz and O. J. F. Martin, "Efficient isotropic magnetic resonators," *Appl. Phys. Lett.*, vol. 81, no. 5, pp. 939-941, 29 July 2002.

[56] P. Winsemius, F. F. van Kampen, H. P. Lengkeek, and C. G. van Went, "Temperature dependence of the optical properties of Au, Ag and Cu," *Journal of Physics F: Metal Physics*, vol.6, no.8, Aug. 1976, pp.1583-606.

[57]  P. T. B. Shaffer, "Refractive index, dispersion, and birefringence of silicon carbide polytypes," *Appl. Opt.*, Vol. 10, No. 5, May 1971, pp. 1034-1036.